\documentclass[12pt,preprint]{aastex}

\newcommand{\lapprox}{{\footnotesize $\buildrel < \over \sim \,$}}
\newcommand{\gapprox}{{\footnotesize $\buildrel > \over \sim \,$}}
\newcommand{\rb}[1]{\raisebox{0.75ex}[0pt]{#1}}

\shorttitle{Modeling Extreme Carbon Stars}
\shortauthors{Speck et al.}

\begin{document}


%\title{The effect of stellar evolution on dust grain sizes} 
%\title{Modeling Extreme Carbon Stars} 
%\title{Silicon Carbide Absorption in Extreme Carbon Stars: 
%Radiative Transfer Modeling}
\title{Silicon carbide absorption features: 
dust formation in the outflows of extreme carbon stars}


\author{ Angela K. Speck, Adrian B. Corman,  
Kristina Wakeman \& Caleb H. Wheeler}
\affil{Department of Physics \& Astronomy, University of Missouri, Columbia, 
MO 65211}
\email{speckan@missouri.edu}

\and
\author{Grant Thompson}
\affil{University of Kentucky, 600 Rose Street, Lexington, Kentucky 40506}


\begin{abstract}

Infrared carbon stars without visible counterparts are generally known as 
extreme carbon stars. We have selected a subset of these stars 
with absorption features in the 10-13\,$\mu$m range, which has been 
tentatively attributed to silicon carbide (SiC). We add three new objects 
meeting these criterion to the seven previously known, bringing our total 
sample to ten sources. We also present the result of radiative transfer 
modeling for these stars, comparing these results to those of previous 
studies. In order to constrain model parameters, we use published mass-loss 
rates, expansion velocities and theoretical dust condensation models 
to determine the dust condensation temperature. These show 
that the inner dust temperatures of the dust shells for these sources are 
significantly higher than previously assumed. This also implies that the 
dominant dust species should be graphite instead of amorphous carbon. In 
combination with the higher condensation temperature we show that this results 
in a much higher acceleration of the dust grains than would be expected from 
previous work.  
Our model results suggest that the very optically thick stage of evolution  
does not coincide with the timescales for the superwind, but rather, that 
this is a very short-lived phase.
Additionally, we compare model and observational 
parameters in an attempt to find any correlations. Finally, we show that the 
spectrum of one source, IRAS 17534$-$3030, strongly implies that the 
10-13\,$\mu$m feature is due to a solid state rather than a molecular species.

\end{abstract}


%% subject header.  Objects should be in the appropriate "individual"
%% headers (e.g. quasars: individual, stars: individual, etc.) with the
%% additional provision that the total number of headers, including each
%% individual object, not exceed six.  The \objectname{} macro, and its
%% alias \object{}, is used to mark each object.  The macro takes the object
%% name as its primary argument.  This name will appear in the paper
%% and serve as the link's anchor in the electronic edition if the name
%% is recognized by the data centers.  The macro also takes an optional
%% argument in parentheses in cases where the data center identification
%% differs from what is to be printed in the paper.



\section{Introduction}

\subsection{Physical evolution of intermediate mass stars and their circumstellar shells}

Stars between about 1 and 8\,M$_{\odot}$ will eventually evolve up the 
Asymptotic Giant Branch  \citep[AGB;][]{Iben1983}. Because of instabilities in 
their interior, AGB stars pulsate and throw off large amounts of mass from 
their surface 
\citep[e.g.,][]{Vassiliadis1993}. This intensive mass loss produces a circumstellar shell of dust and neutral gas.
Once the AGB star has exhausted its outer envelope, the AGB phase ends.
At this point, the mass loss virtually stops and the circumstellar shell
begins to drift away from the star.
At the same time, the central star begins to
shrink and heat up from $\sim$3000\,K until it is hot enough to ionize the
surrounding gas, at which point the object becomes a planetary nebula (PN).
The short-lived post-AGB phase, as the star evolves toward to the PN phase, is
also known as the proto- or pre-planetary nebula (PPN) phase.
 During the ascent of the AGB, the
velocity of the outflowing mass appears to be fairly constant
\citep[e.g.,][]{hug88,fong03}. Therefore the dust furthest from the star
represents the oldest mass loss, while material closer to the star represents
more recent mass loss. Towards the end of the AGB phase
the increasing
 impact of the thermal pulse cycles leads to an increasing mass-loss
 rate 
\citep[e.g.,][]{vw93,vil02a}. Such an increase in mass-loss rate (dubbed the
{\it superwind}) is necessary to explain the densities seen in typical PNe 
\citep{renz81}. Since the invocation of the superwind,
many observations of AGB stars and post-AGB stars have supported this 
hypothesis \citep[e.g.,][]{km85,wood92}.

\subsection{Chemical evolution of AGB stars and their circumstellar shells}

The chemical composition of the atmospheres of AGB stars is expected 
to change as these stars evolve, due to convective dredge up of carbon 
produced in the He-burning shell. The amount of carbon relative to oxygen (the 
C/O ratio) is critical in determining which types of dust and molecules 
are present around 
an AGB star. The formation of extremely stable CO molecules will consume 
whichever of the two elements is less abundant, leaving only the more abundant 
element available for dust formation. Stars start their lives with the cosmic 
C/O ratio of $\approx$0.4 and are therefore oxygen-rich. In about a third of 
AGB stars, enough carbon will be dredged up to make C/O $>$ 1 and therefore 
carbon will dominate the chemistry around these stars, known as carbon stars.
Carbon stars are expected to have circumstellar shells dominated by amorphous 
or graphitic carbon grains, which do not have diagnostic infrared features. 
Another component of the dust shell around carbon stars, silicon carbide (SiC),
does have an infrared spectral feature at $\approx$11$\mu$m and therefore has 
been of great interest to researchers seeking to understand the evolution of 
the dust shells and infrared features of carbon stars 
\citep{Baron1987, Chan1990, Goebel1995, Speck1997, Sloan1998, Speck2005, 
Speck2006, Thompson2006}. 

\subsection{Extreme Carbon Stars}
\label{ecs_intro}

As carbon stars evolve, mass loss is expected to increase. Consequently, their 
circumstellar shells become progressively more optically thick, and
eventually the central star is obscured. \citet{Volk1992,Volk2000} 
christened such stars ``extreme carbon  stars''. 
These stars have also been dubbed ``infrared carbon stars'' 
\citep{Groenewegen1994}, and ``very cold carbon stars'' \citep{omont93}.
Extreme carbon stars are expected represent that small subset of 
carbon-rich AGB stars which are in the superwind phase, just prior to 
leaving the AGB. 
Because the superwind phase is short-lived compared to the AGB phase the 
number of extreme carbon stars is intrinsically small. Consequently, few 
of these objects are known. At present there are $\sim$30 known extreme carbon stars in the Galaxy \citep{Volk1992}
compared to $\sim$30,000 known visible carbon stars \citep{skrutskie}. 

 \citet{vh88} attempted to define a way to distinguish between oxygen-rich 
and 
carbon-rich AGB stars using IRAS color-color space, which was divided into 
subsections according to the properties of the dusty shells are these stars 
\citep[see Table~1 of][]{vh88}. This was further refined 
by \citet{omont93} who identified a population of very cold carbon stars using 
HCN and CO observations, and showed that the regions originally designated as 
extremely dusty O-rich AGB stars also contain a significant fraction of C-rich 
stars.

\begin{figure}[t]
%\resizebox{\hsize}{!}{%
\includegraphics[angle=270,scale=0.75]{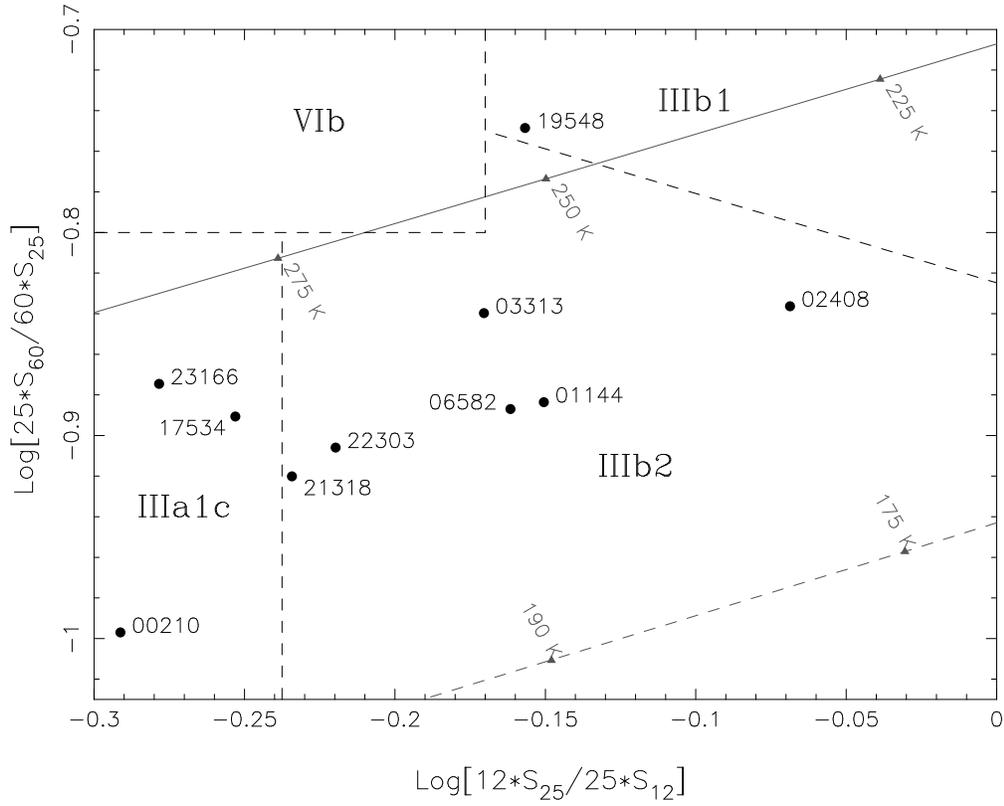}%}
\caption{Regions in the IRAS color-color diagram populated by extreme carbon stars. see text for details. 
The black body curve is indicated by the solid grey line.
The modified blackbody [$B(T,\lambda)*\lambda^{-1}$] is indicated by the dashed grey line}
\label{irascolor} 
\end{figure}

The refinement of the \citet{vh88} color-color diagram by \citet{omont93} 
defined subdivisions of the seven zones in color-color space 
\citep[see Fig.~1 in][]{omont93}.
Cool carbon stars with high mass-loss rates (and little or no SiC emission) 
fall into regions III and IV, which had previously been assumed to define 
OH-IR stars (i.e. the oxygen-rich counterparts to extreme carbon stars). The 
numbered regions have been subdivided into smaller regions denoted by 
IIIa1, IIIa2, IIIb1 etc.
%
%{\tt [ need to say what the subdivisions mean]} 
%
A subset of the color-color space, covering parts of
regions IIIa1c, IIIb2, IIIb2 and VIb is reproduced in Fig.~\ref{irascolor} and 
includes our sample stars.

\subsection{SiC absorption features}
\label{sicabs}

SiC has long been predicted to be present in carbon star circumstellar shells, 
beginning with condensation theories \citep{Friedman1969,Gilman1969} and 
continuing with the prediction of a characteristic SiC $\sim$11$\mu$m  
spectral feature \citep{Gilra1971} and then the observational discovery of 
an $\sim$11$\mu$m emission feature in many carbon star spectra 
\citep{Hackwell1972,Treffers1974}. 
The effect of the evolving dust shell density structure on observed 
features, and particularly on the $\sim$11$\mu$m feature, have been discussed 
extensively 
\citep[see review in][and references therein]{Speck2005}.
As the optical depth of the dust shell increases, self-absorption
will diminish the $\sim$11$\mu$m feature and it will 
eventually be seen in net absorption. 
SiC self-absorption was found to be important in producing accurate 
radiative transfer models of extreme carbon stars
\citep[e.g.][]{Volk1992}, even though this previous work did not recognize
SiC absorption features.
These absorption features are rare and have mostly been ignored in 
discussions of 
evolutionary sequences in carbon star spectra. In fact the rarity of such 
absorption features led to the hypothesis that SiC becomes coated in carbon at 
high optical depths \citep[e.g.][]{Baron1987,Chan1990}. 
However, meteoritic data and theoretical models do not support this hypothesis 
(see \S~\ref{meteor} and \S~\ref{condmod}).

A few extreme carbon stars have been shown to have an absorption feature at 
$\sim11\mu$m which has been tentatively attributed to SiC. This feature was 
discovered in the ``prototype'' extreme carbon star AFGL~3068 
\citep[hereafter referred to as IRAS\,23166$+$1655;][]{Jones1978}, 
and was re-examined by \citet{Speck1997}, which also 
identified three additional extreme carbon stars with this feature. 
\citet{Clement2003} examined the absorption features of two of these extreme 
carbon stars (IRAS\,23166+1655 and IRAS\,02408+5458), 
and showed that their 11$\mu$m 
absorption features are consistent with $\beta$-SiC\footnote{%
Silicon carbide exists in many ($>$70) different crystal structures, known as 
polytypes. See \citet{Speck1997,daulton03,Pitman2007} for a discussion of 
the polytypes of SiC.}
nanoparticles. 
The broad absorption features of IRAS\,19548+3035 and IRAS\,21318+5631
\citep[also discovered by][]{Speck1997} were attributed to SiC absorption
with an interstellar silicate absorption contribution 
\citep[see also][]{Groenewegen1996}. This will be discussed further in 
\S~\ref{correlsect}.
The absorption features in the spectra of IRAS\,19548+3035 and IRAS\,21318+5631
were revisited by \citet{Clement2005} who suggested Si$_3$N$_4$ grains as the 
carrier. However, this hypothesis has been shown to be erroneous
\citep{Pitman2006}. 

The failure of the Si$_3$N$_4$ hypothesis led \citet{Speck2005} to suggest that
amorphous SiC grains may be able to account for the breadth, structure  and 
barycentric position of the observed broad 10-13$\mu$m feature in 
IRAS\,19548+3035 and IRAS\,21318+5631. 
However, the dearth of amorphous presolar SiC grains seems to 
preclude this hypothesis (see \S~\ref{meteor}).
An alternative explanation for this feature is molecular line absorption, 
however, currently available line lists are not sufficient to properly 
assess this hypothesis \citep[see][and references therein]{Speck2006}.
One molecular candidate which has transitions in the correct wavelength 
range is C$_3$
\citep[e.g.][]{zijl06,jorg}, but the line lists are not readily 
available. Furthermore, C$_3$ is expected to be photospheric, rather than circumstellar, 
which probably precludes its detection in optically obscured stars.
Moreover, the theoretical spectrum of C$_3$ 
from \citet{jorg} shows a strong absorption close to the $\sim5\mu$m CO 
line, which is stronger than the $\sim11\mu$m feature. As will be seen in
\S~\ref{17534}, the spectrum of IRAS\,17534$-$3030 does not show the 5$\mu$m 
absorption band and provides evidence that the
observed absorption feature is not molecular in origin.

Though previous research has included the effects of SiC self-absorption 
\citep[shown to be crucial to produce accurate 
models][]{Volk1992,Speck1997,Speck2005}, 
no work has been done to directly fit the apparent SiC absorption feature in 
radiative transfer models of extreme carbon stars.

\subsection{Previous Radiative Transfer Models of Extreme Carbon Stars}
\label{rtmodels1}

\citet{Volk1992} performed radiative transfer modeling in order to match the
Infrared Astronomical Satellite \citep[{\em IRAS};][]{neug84} Low Resolution 
Spectrometer (LRS) data for several of extreme carbon stars.
They determined that the exact star temperature entered into the 
model was not important for the emerging spectra due to the very thick dust 
shells around extreme carbon stars 
\citep[c.f.][]{DePew2006,Speck2000}.
Their models used a fixed composition (a mixture of graphite 
and SiC), and a fixed dust condensation temperature. 
\citet{Groenewegen1994} also performed radiative transfer modeling on a larger 
set of extreme carbon stars, but these models varied the dust condensation 
temperature. Again this was based on {\em IRAS} LRS data.
Following up on this, \citet{Groenewegen1995} modeled a large sample of carbon stars
using amorphous carbon optical constants \citep{rm91}, 
and assumed low dust condensation temperature in a fairly narrow range 
(650--900\,K for the extreme carbon stars). Consequently the inner dust radius is 
larger than expected. Moreover the resulting models all
have relatively low optical depths ($\tau_{11.3\mu \rm m} < 2$). 
The optical depth for their IRAS\,23166+1655 model
was found to be $\tau_{11.3\mu \rm m} < 1$, 
even though this star has an absorption feature at 11$\mu$m.
\citet{Groenewegen1998} remodeled these stars, again assuming relatively low 
dust condensation temperatures, with similar results.

Finally, \citet{Volk2000} used the improved spectral resolution of the 
the Infrared Space Observatory \citep[{\em ISO};][]{kessler} Short Wavelength 
Spectrometer \citep[SWS;][]{degraauw96} to examine five extreme carbon stars. 
In their modeling study, \citet{Volk2000} allowed the optical depth and radial 
dust density distribution to vary; the resulting optical depths were 
relatively high (1.4-4.5 at 11.3$\mu$m), and the density of the dust shell was 
found to increase rapidly towards the center. This increase was interpreted as 
evidence of an increasing mass-loss rate within the last few thousand years, 
consistent with the identification of extreme carbon stars as the final stage 
of AGB star evolution. While these models did include SiC opacity data, the 
11$\mu$m absorption feature was not recognized and consequently no attempt was 
made to fit this feature in these models.

A summary of the parameters of previous models for extreme carbon stars with 
11$\mu$m absorption features in our sample can be 
found in Table~\ref{prevmod}. Interestingly, all previous models assume
relatively low inner dust temperatures. This will be discussed further in 
\S~\ref{condmod}.
Furthermore, the modeled dust density distributions suggest a relatively slow 
increase in mass-loss rates ($1/r^x$, where $x \approx$ 2.25---3.0). 
%This is slower than the predicted increase \citep[e.g.,][]{vw93}.

In reality, dust shells are expected to have heterogeneities and anisotropies 
in their density structure 
as a result of pulsation-driven dust formation and the 
ensuing hydrodynamic turbulent effects \citep[e.g.][]{woitke06}. 
These dust formation models suggest that carbon star mass-loss is expected 
to be modulated on several timescales, especially that of the pulsation cycle. 
Furthermore \citet{woitke06} has suggested that 
the dynamics in the dust-forming zones around carbon stars lead to 
inhomogeneous dust formation, producing fine scale structure in the density of> the dust envelope.
In addition, while pulsation shocks are predicted to have a strong effect on 
local conditions \citep[e.g.][]{cherchneff06}, 
this is not reflected in temporal changes in the IR spectra of carbon stars 
\citep{corman08}.
As will be seen in \S~\ref{results}, the 
spatial scale of the heterogeneities is small and the 
timescale for pulsations is short compared to the timescales associated with 
even the thinnest dust 
shells. Moreover, the inhomogeneities are expected to be wiped out over time 
by the hydrodynamic interactions \citep{vil02a,vil02b}. Consequently, we do 
not consider these small scale structures in our models.

\subsection{Meteoritic Evidence}
\label{meteor}

The isotopic compositions of certain  grains found in primitive meteorites 
indicate that they originated outside the solar system and are thus dubbed 
``presolar''.
Dust grains from AGB stars are found virtually unaltered in
these meteorites, demonstrating that these grains become part of the next
generation of stars and planets \citep[][and references therein]{cn04}.
The precise physical characteristics of these meteoritic dust grains
(e.g. sizes, crystal structures, compositions) can be used to help constrain
the nature of the dust we see in our astronomical observations.

\subsubsection{Presolar silicon carbide}
\label{presolsic}

Silicon carbide was the first presolar grain to be found in meteorites
\citep{Bernatowicz1987} and remains the best studied 
\citep[][and references therein]{bern05}. 
The most important findings of this work are 
(1) that most ($\sim99$\%) of the SiC presolar grains were formed around 
carbon stars; 
(2) of the AGB SiC grains, \gapprox 95\% appear to originate around low-mass 
carbon stars ($<$3\,M$_\odot$), based on nucleosynthesis models of isotopic compositions;
(3) that all the SiC grains are crystalline (not amorphous);
(4) that nearly all (\gapprox80\%) are of the cubic $\beta$ polytype, with the 
remainder comprising the lower temperature 2H polytype;
(5) that with one exception, SiC grains have not been 
found in the cores of carbon 
presolar grains (unlike other carbides: TiC, ZrC, and MoC); and 
(6) that the grain size distribution includes both very small and very large 
grains (1.5\,nm $\rightarrow$ 26\,$\mu$m), with most grains in the 
0.1--1$\mu$m range.
Single-crystal grains can exceed 20$\mu$m in size.
Observations of the 11$\mu$m feature have been compared with
laboratory spectra of various forms of SiC, and after some false
starts it has now been attributed to $\beta$-SiC, matching the information
retrieved from meteoritic samples \citep{Speck1999,Clement2003}. 
However, there are still some discrepancies between
observational and meteoritic evidence (most notably related to
grain size).

 \citet{prombo}
found a correlation between grain size and the concentration of 
{\it s-process} elements in SiC grains taken from the Murchison meteorites. 
The Indarch meteorite presolar SiC grains yielded similar results 
\citep{jennings}. In both cases, the smaller grains have higher relative 
abundances of {\it s-process} elements. This observation may be a result of 
different metallicity sources yielding different grain-size distributions
\citep{Lagadec2007,Lagadec2008}. Alternatively, it may reflect an evolution 
in grain-size with dredge-up \citep{Speck2005}.

\subsubsection{Presolar ``graphite''}
\label{presolc}

In addition to SiC presolar grains, carbon grains are also relatively abundant 
and well studied \citep[see][and references therein]{bern05}.
Presolar carbon grains are usually referred to as ``graphite'' grains, but 
their structures are more complex than this name infers.
Presolar graphite is found in two types of spherules classified according to 
their external morphologies as ``onion-like'' and ``cauliflower-like''.
In general the graphite spherules follow a similar size distribution to the 
SiC grains. However, the high-density grains 
($\rho \approx 2.15-2.20$\,g\,$cm^3$) associated with AGB stars have a 
mean size of 2$\mu$m. In addition, the AGB presolar graphite spherules span a 
larger range of isotopic compositions than the SiC grains, possibly suggesting 
that they form at a wider range of times during the AGB phase.

While the presolar SiC grains tend to be single crystals, the graphite grains 
regularly contain carbide grains. These carbides are enriched in $s$-process 
elements, indicative of formation around late-stage AGB stars.
Many of the ``onion-like'' graphite grains have a core mantle structure in which the 
core contains disordered agglomerations of graphene\footnote{%
Graphene is basically a single sheet of graphitic material. If it is 
disordered, there are some heptagons and pentagons in place of the regular 
hexagonal carbon structure. Graphite is the 3-d structure.}
sheets and PAH\footnote{polycyclic aromatic hydrocarbon}-like 
products, while the mantle is composed of well-ordered graphitic concentric 
shells. The graphene particles have a typical size of 3-4nm.
The ``cauliflower-like'' graphite grains also have a concentric shell 
structure, but it is less well ordered, and is composed primarily of the 
disordered graphene.
Whether ``onion'' and ``cauliflower'' graphites are formed in the same 
outflows is not known. Both types of grain contain the refractory carbides and 
both span the same range of isotopic compositions. 
Whether the ``onion'' or ``cauliflower'' grains are more representative of 
grains in the outflows of extreme carbon stars is not known. However, even the 
most disordered ``cauliflowers'' or ``onion''-cores are still closer to 
graphite 
than glassy carbon in structure. The least ordered grains are still considered 
to be agglommerations of nano-crystalline grains, rather than truly amorphous
(pers.\ comm.\ K. Croat).

\subsubsection{Other presolar carbides}
\label{presoltic}

As discussed in \S~\ref{presolsic} and \S~\ref{presolc}, refractory carbides 
are found inside ``graphite'' grains but not in SiC grains. Furthermore, SiC 
is not one of the carbides found in ``graphite'' grains. The refractory 
carbides (TiC, ZrC, MoC and RuC) provide more constraints on the dust 
formation processes around carbon stars. In particular, the formation of 
``graphite'' spherules with TiC nuclei limits the range of C/O ratios in which 
these grains could form to 1 \lapprox C/O \lapprox 1.2.
Meanwhile, the ZrC can form nuclei at higher C/O, but the value still needs to 
be less than two.
This is consistent with the measured C/O ratios of Galactic carbon stars, 
which have an average of 1.15 and a maximum of 1.8 \citep{lamb86}

\subsection{Investigation}
In the present work, we investigate a subset of extreme carbon stars, those 
which exhibit the 11$\mu$m absorption feature. Through radiative transfer 
modeling, we investigate the nature of these dust shells. We use theoretical 
models and meteoritic data to limit the parameter space and thus reduce the 
degeneracy within the model results. 
In addition, we look for correlations between observed parameters, such as 
those that define the 11$\mu$m feature (strength, position, etc) as well as 
mass-loss rates and expansion velocities associated with the dust shells.
Finally we determine timescales associated with the dust shells.

\section{Observations and Data Processing}
\label{obssect}

\begin{figure}[t]
%\resizebox{\hsize}{!}{%
\includegraphics[angle=270,scale=0.85]{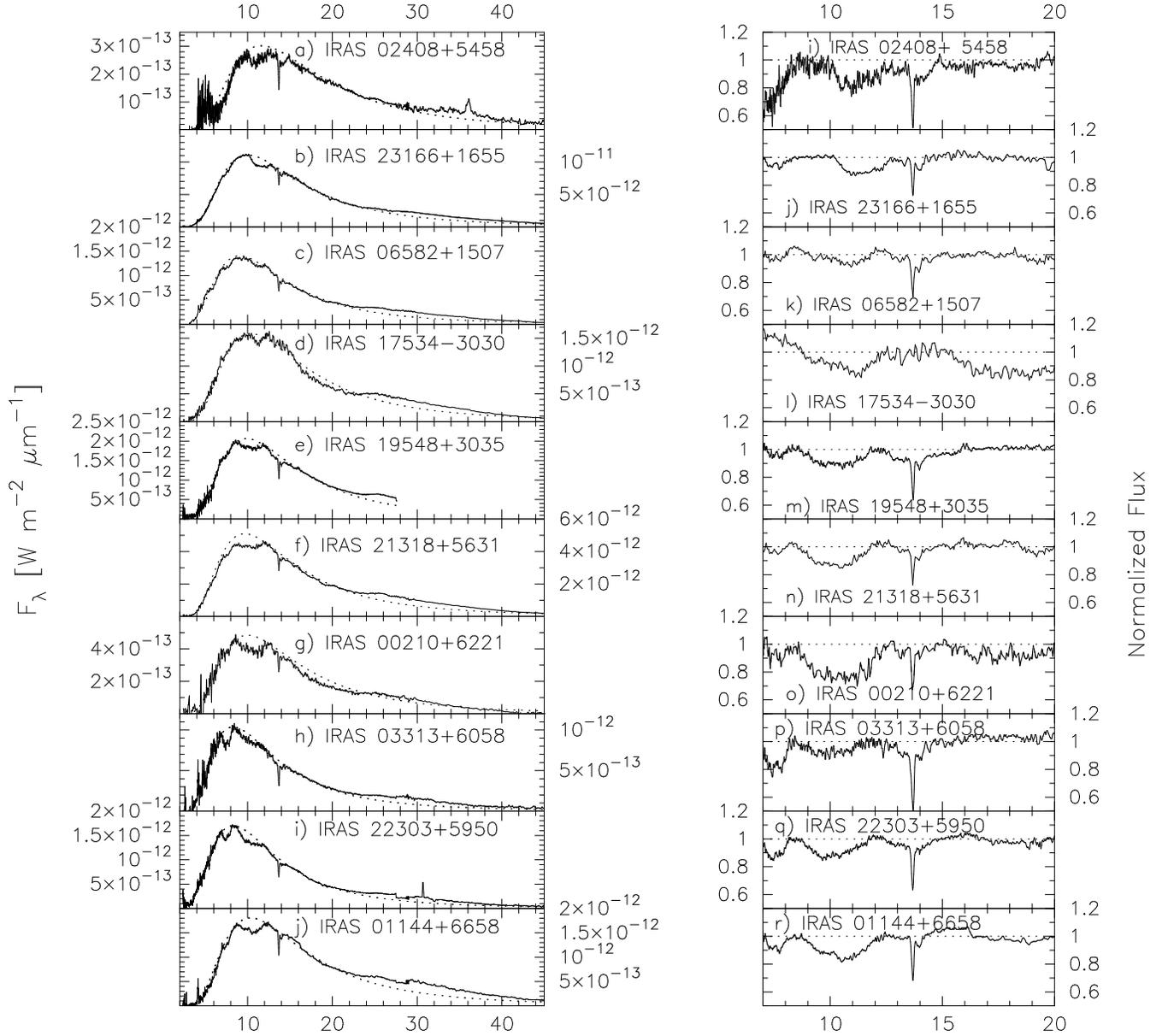}
\caption{{\it ISO\/} SWS spectra of ten extreme C-stars.
{\it Left Panel\/}: Flux-calibrated spectra and fitted blackbody continua.
Solid line = {\it ISO\/} spectrum;
dashed line = best-fitting blackbody continuum;
The $y$-axis is the flux (F$_\lambda$) in W\,m$^{-2}$\,$\mu$m$^{-1}$;
the $x$-axis is wavelength in $\mu$m.
{\it Right Panel\/}: Continuum-divided spectra.
%Solid line = {\it ISO\/} spectra; dashed line = ground-based spectra from^M
%Speck et al. (1997). 
Table~\ref{obsparam} lists the blackbody temperatures used in each case to 
produce the continuum and the continuum-divided spectra.}
\label{obsfig} 
\end{figure}

\begin{figure}[t]
%\resizebox{\hsize}{!}{%
\includegraphics[angle=270,scale=0.7]{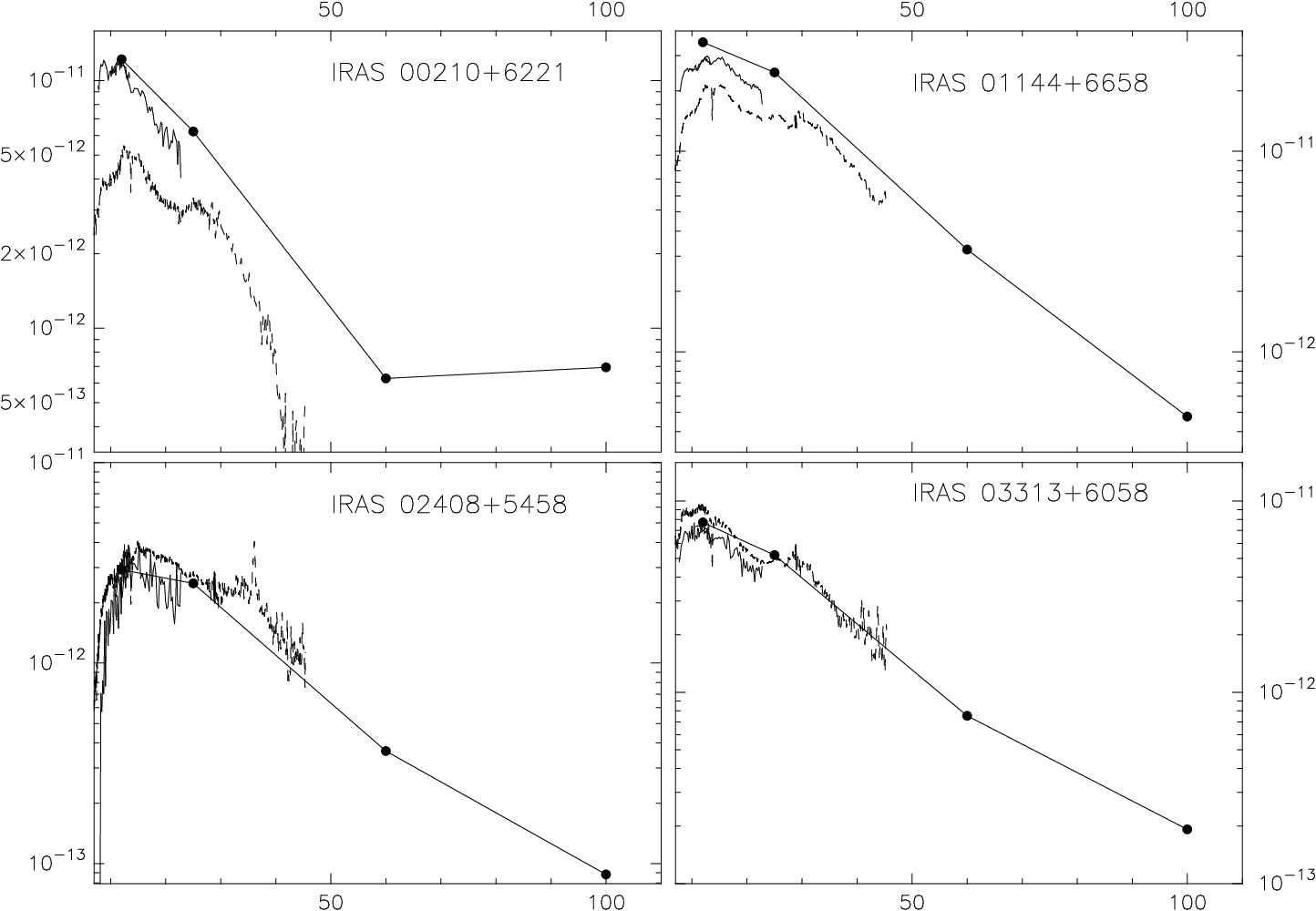}
\caption{{\it ISO\/} SWS spectra of sample extreme C-stars together with 
{\it IRAS\/} 12, 25, 60 and 100$\mu$m photometry points and {\it IRAS\/} LRS 
spectra.
Solid line = {\it IRAS\/} spectrum;
points =  {\it IRAS\/} photometry points;
dashed line = {\it ISO\/} SWS spectrum.
The $y$-axis is the flux ($\lambda$F$_\lambda$) in W\,m$^{-2}$;
the $x$-axis is wavelength in $\mu$m.}
\label{obsfig2} 
\end{figure}

\begin{figure}[t]
%\resizebox{\hsize}{!}{%
%\addtocounter{figure}{-1}
\includegraphics[angle=270,scale=0.7]{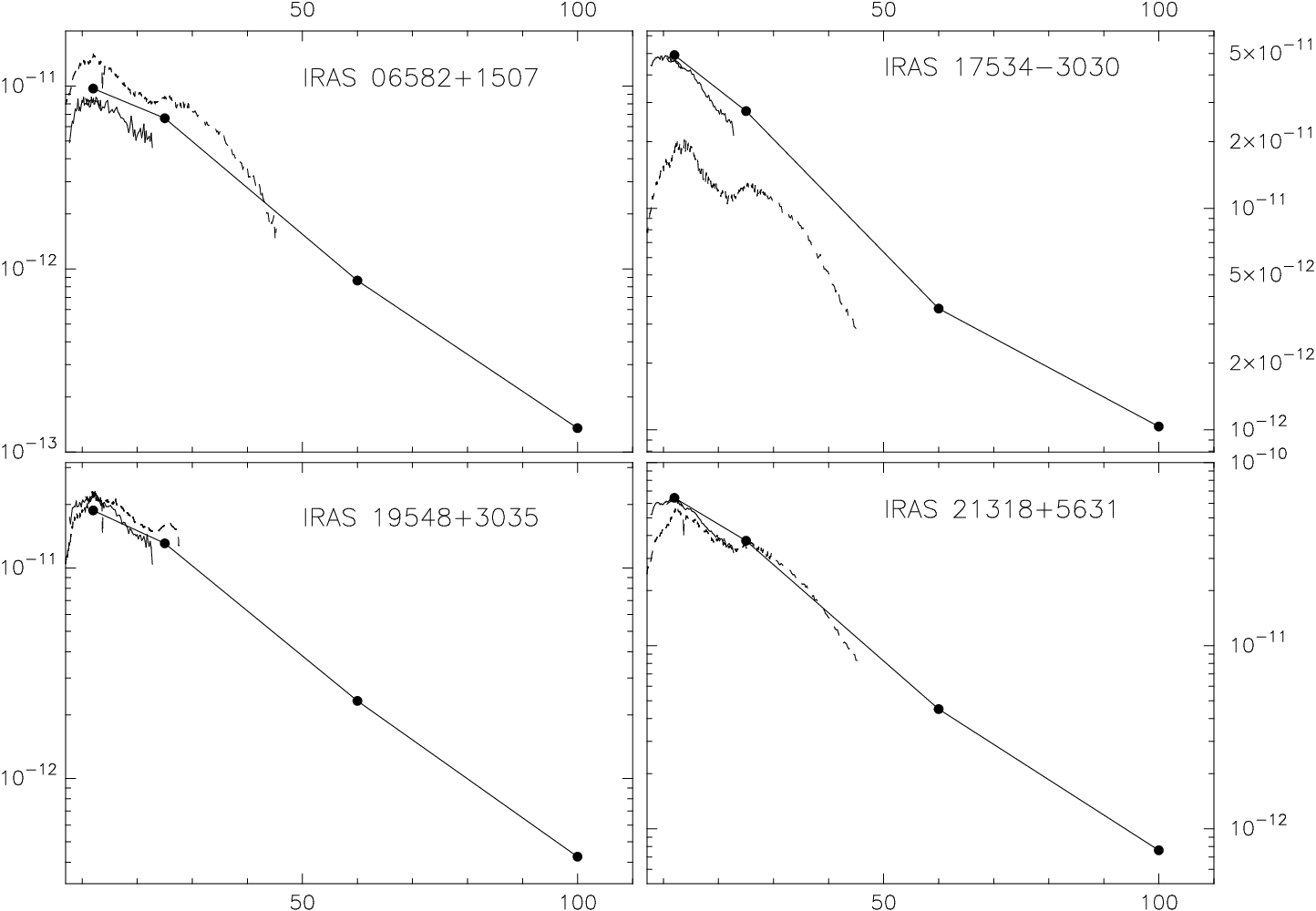}
\caption{{\it ISO\/} SWS spectra of sample extreme C-stars together with 
{\it IRAS\/} 12, 25, 60 and 100$\mu$m photometry points and {\it IRAS\/} LRS 
spectra (part 2).
Solid line = {\it IRAS\/} spectrum;
points =  {\it IRAS\/} photometry points;
dashed line = {\it ISO\/} SWS spectrum.
The $y$-axis is the flux ($\lambda$F$_\lambda$) in W\,m$^{-2}$;
the $x$-axis is wavelength in $\mu$m.}
\label{obsfig3} 
\end{figure}

\begin{figure}[t]
%\resizebox{\hsize}{!}{%
%\addtocounter{figure}{-1}
\includegraphics[angle=270,scale=0.7]{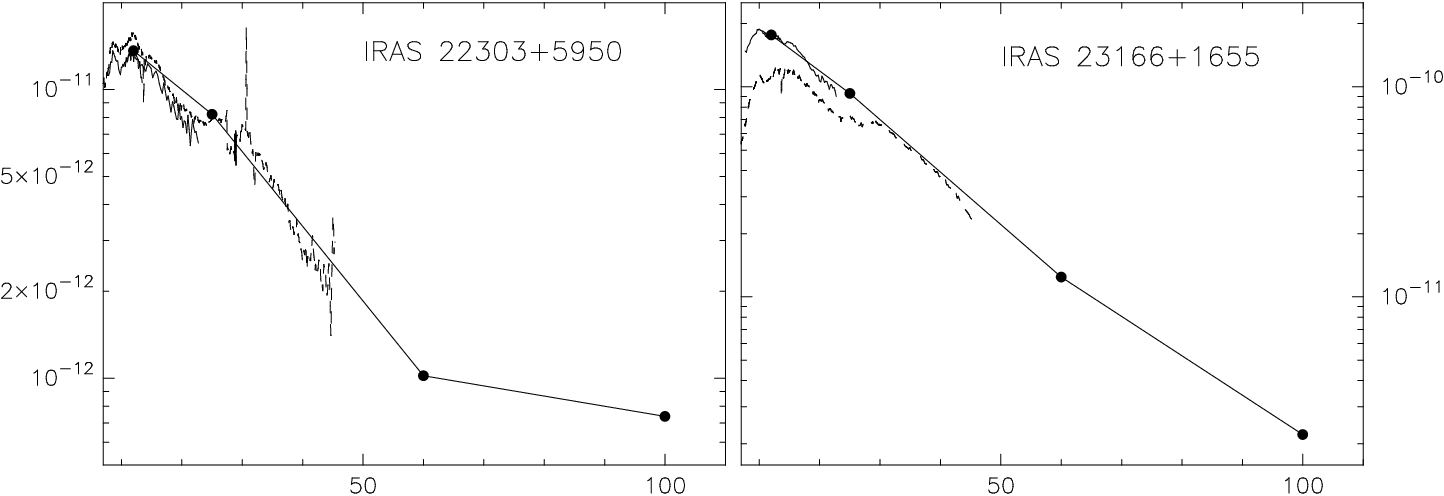}
\caption{{\it ISO\/} SWS spectra of sample extreme C-stars together with 
{\it IRAS\/} 12, 25, 60 and 100$\mu$m photometry points and {\it IRAS\/} LRS 
spectra (part 3).
Solid line = {\it IRAS\/} spectrum;
points =  {\it IRAS\/} photometry points;
dashed line = {\it ISO\/} SWS spectrum.
The $y$-axis is the flux ($\lambda$F$_\lambda$) in W\,m$^{-2}$;
the $x$-axis is wavelength in $\mu$m.}
\label{obsfig4} 
\end{figure}

We investigated 10 extreme carbon stars observed spectroscopically by the ISO 
SWS all of which show evidence for an $\sim11\mu$m absorption feature 
(see Table~\ref{obstable} and Figs.~\ref{obsfig}, \ref{obsfig2},\ref{obsfig3}
and \ref{obsfig4}). 
These sources were chosen by searching the ISO archive 
for spectra of extreme carbon stars listed in \citet{Volk1992} and selecting 
those with an apparent 11$\mu$m absorption feature. In addition, we used the 
color-color classification of ``very cold'' carbon stars by \citet{omont93} to 
identify further potential sources. Unfortunately most of the potential 
sources found in the color-color space (e.g. IRAS\,17583-2291) were not observed
by ISO SWS, and the IRAS LRS spectra are too low resolution and/or too noisy to
 be used in the present study.
Four of our sources (IRAS\,02408+5458, IRAS\,19548+3035, IRAS\,21318+5631, and 
IRAS\,23166 +1655) 
were previously studied using ground-based observations and were found to be 
consistent with a self-absorbed SiC feature  \citep{Speck1997}. 
Two of these sources needed an extra absorption component at $\sim$10$\mu$m, 
which were attributed to interstellar absorption. As discussed in 
\S~\ref{sicabs}, these have since been the source of some controversy 
\citep{Clement2005,Pitman2006}. 
Following the modeling efforts of \citet{Volk2000}, \citet{Speck2005} 
identified IRAS\,00210+6221, IRAS\,06582+1507, and IRAS\,17534$-$3030 as exhibiting 
an SiC absorption feature. The modeling of \citet{Volk2000} did not include 
SiC and did not attempt to fit the 11$\mu$m absorption. Thus, division of the 
observed spectra by their respective RT model spectra revealed the 11$\mu$m 
absorption feature. 
In this paper we present the discovery of three more potential SiC absorption 
features in ISO SWS spectra (IRAS\,01144+6658, IRAS\,03313+6058, and 
IRAS\,22303+5950). These were discovered by searching the ISO archive for any 
extreme or ``very cold'' carbon stars as determined by their location in the 
IRAS color-color space. Those sources that fall within region III without 
OH maser emission were examined.
The locations of the stars in our sample in {\em IRAS} color-color space
\citep{vh88,omont93} are plotted in Fig~\ref{irascolor}.
Interestingly, all sources except IRAS\,19548+3035 
plot along a line parallel to and between the 
blackbody emission and the $B(T,\lambda,T)\times\lambda^{-1}$ emission lines.

The raw ISO data were extracted from the ISO data archive, and we used the 
Off-Line Processing (OLP) pipeline, version 10.1. Individual spectral 
sub-bands were cleaned of glitches (caused by cosmic ray particles) and other 
bad data sections. Next, they were flat-fielded, sigma-clipped (using the 
default values $\sigma$ = 3) and rebinned to the final spectral resolution 
($R=\Delta\lambda/\lambda$), which ranged from 200 to 700, depending on the scanning 
speed of the SWS grating during the observation \citep{Leech2003}. The final 
spectra are presented in Figure~\ref{obsfig}, which also shows the 
best-fitting\footnote{%
Best fits are achieved by eye and proceed by examination of the 
continuum-divided spectra.}
blackbody continuum for each spectrum and the resulting 
continuum-divided spectra. The blackbody temperatures of the continua at 
listed in 
Table~\ref{obsparam}.
The continuum-divided spectra clearly show an absorption feature in the 
10-13$\mu$m region, the basic parameters of which (barycentric position, 
peak-to-continuum ratio, full width half maximum; FWHM) are listed in 
Table~\ref{obsparam}.
The excellent match between the overall spectrum and a single temperature 
blackbody suggests that we are seeing an isothermal surface within the dust 
shell. This represents the depth at which the shell becomes optically thick. 
The lack of extra emission at longer wavelengths suggests that any outlying 
dust is low enough in density to have an insignificant contribution to the 
overall emission. 
%This, in turn, suggests that we are only seeing the {\em superwind}, 
%$and the mass-loss rate prior to the onset of the superwind was low.
%This will be discussed further in \S~\ref{timescalesect}.

In addition to the {\it ISO\/} SWS spectra, Fig~\ref{obsfig2}, \ref{obsfig3}
and \ref{obsfig4} shows the 
{\it IRAS\/} LRS spectra and the {\it IRAS\/} 12, 25, 60 and 100$\mu$m 
photometry measurements. The difference in the flux levels between the
{\it IRAS\/} and  {\it ISO\/} data is not unexpected, since these stars are 
variable. However, the shape of the spectrum does not change significantly 
between the two observations, suggesting that changes in the stellar 
luminosity do not significantly impact the structure and composition 
of the dust shells.

In order to determine the cause of the 10-13$\mu$m feature and the factors 
that govern its strength, width and position we have tabulated the barycentric 
position, feature-to-continuum ratio and equivalent width of the feature 
(Table~\ref{obsparam}). In addition, we have also tabulated where the 
barycenter of the SiC feature would be if the short wavelength side of the 
absorption is due to silicate (as has been postulated, see \S~\ref{sicabs}).
This, along with the feature-to-continuum ratio measured at 9.7 and 
11.3$\mu$m, can be found in Table~\ref{obsparam}.

\subsection{The ``30$\mu$m'' feature}

Another prominent spectral feature exhibited by our sample of extreme carbon stars is the  so-called ``30$\mu$m'' feature.
This feature is relatively common amongst carbon-rich AGB stars, PPNe 
and PNe and  was first discovered in Kuiper Airborne Observatory ({\it KAO}) 
spectra of carbon stars and PNe \citep{forrest81}. It has been widely 
attributed to magnesium sulfide \citep[MgS; e.g.][]{gm85,nuth85,omont95,begemann94,hony2002}.
Modeling this feature is beyond the scope of the present work, but will be 
investigated in follow-up modeling. Our models make no attempt to fit the ``30$\mu$m'' feature.
%spectrum longwards of $\sim26\mu$m.

\section{Radiative Transfer Modeling}
\label{rtmodeling}

Radiative transfer modeling has been particularly useful in investigating 
extreme carbon stars 
\citep[see \S~\ref{rtmodels1};][]{Volk1992,Groenewegen1995,Volk2000}. 
We used the 1-D radiative transfer program DUSTY 
\citep{Ivezic1995, Nenkova2000}, to determine the effect of dust shell 
parameters on the emerging spectra from carbon stars. 
In all cases, the central star was assumed to be at 3000\,K (typical 
for an AGB star). \citet{Speck2000} and \citet{DePew2006} showed that 
changing this 
temperature by $\pm$1000\,K did not significantly change the radiative 
transfer model's spectra \citep[c.f.][]{Volk2000}. For simplicity, 
dust grains are assumed to be spherical. While DUSTY can include other grain 
shapes, this expands parameter space to create more degeneracy between models 
and is beyond the scope of the present work. 

\subsection{Radial dust density distribution}
\label{radialdist}

We assume a radial dust density distribution of 
$1/r^2$ which would reflect a constant mass-loss rate.
This choice is somewhat controversial and certainly needs justifying.
Current dust formation models suggest that carbon star mass-loss is expected 
to be modulated on several timescales, especially that of the pulsation cycle 
\citep[][and references therein]{woitke06}. Furthermore \citet{woitke06} has 
suggested that 
the dynamics in the dust-forming zones around carbon stars lead to 
inhomogeneous dust formation, producing fine scale structure in the density of 
the dust envelope.
While pulsation shocks are predicted to have a strong effect on local 
conditions \citep[e.g.][]{cherchneff06}, 
this is not reflected in temporal changes in the IR spectra of carbon stars 
\citep{corman08}.
Previous models of extreme carbon stars have included steeper a density 
drop-off \citep[see 
Table~\ref{prevmod};][]{Volk1992,Volk2000,Groenewegen1995,Groenewegen1998},
which is meant to represent the increasing mass-loss rate associated with the
onset of the superwind phase. However, \citet{vil02a,vil02b} 
showed that 
the hydrodynamics in the circumstellar shell wipe out density structure and 
leave a shell with a $1/r^2$ density distribution. In addition 
\citet{rrh83} showed that carbon star spectra can be well-fitted using such a 
density distribution. 

In our models we assume that modulations in density have 
been wiped out or are unimportant in determining the spectrum at these high 
optical depths, as we are clearly seeing an outer dust shell surface.
We also assume we are only detecting the dust that has formed since 
the onset of the superwind. We assume that there was a 
sudden increase in mass loss at some time in the last $\sim$10000\,years 
and that this mass-loss rate is now approximately constant, with small scale 
fluctuations being unimportant for dust properties.
Consequently, the $1/r^2$ density distribution suffices.
As will be seen in \S~\ref{results}, the impact of assuming a steeper the 
density distribution is to remove dependence on shell size, and thus remove 
the ability to out limits on timescales. If we were to adopt a  
$1/r^3$ density distribution, all the dust would effectively be 
contained close to the star and would reflect the total mass lost over only a 
relatively short period.

\subsection{Modeling grain-size distributions}
\label{gszmodel}

The issue of choosing a grain size distribution is interesting and certainly 
requires more discussion. Our initial modeling studies used an MRN distribution
 \citep[i.e., $n(a)$ proportional to $a^{-q}$, where $n$ is the number of the 
grains in the size interval $(a,a + da)$ and $q=3.5$;$a_{min}=0.005\mu$m; and 
$a_{max}=0.25\mu$m;][]{Mathis1977}. 
This was chosen because as will become evident below, 
we do not actually know the grain size 
distribution and MRN is as plausible as any other. 
However, the MRN distribution was developed for interstellar dust where the 
balance of formation and destruction is different from AGB circumstellar 
environments. \citet{dsg89} suggested that the grain-size
distribution created in the circumstellar environments of AGB stars has a 
steeper power law (i.e.\ $a^5$), while \citet[][KMH]{kmh} modified the MRN 
distribution to include an exponential fall-off term.

We should also consider the ``observational'' evidence for range and distribution 
of grain sizes in carbon star outflows. 
Meteoritic presolar SiC grains from carbon stars have a huge grain-size 
distribution, ranging for 1.5nm up to 26$\mu$m, with the majority of grains 
($\sim$70\%) falling in the 0.3--0.7$\mu$m range 
\citep[see][]{daulton03,bern05,cn04}. 
Half the mass of the presolar SiC found in the Murchison meteorite is in 
grains larger than 0.6$\mu$m \citep{virag}. 
Carbon presolar grains follow a similar grain size distribution to SiC 
\citep{bern05}.
However, the sample may be biased 
in favor of large grains, which may be more apt to survive the journey through 
the ISM and incorporation into a solar system body. 

Conventional theories of 
grain growth cannot produce the largest grains. Since AGB stars typically have 
mass-loss rates $ < 10^{-5}$ M$_\odot$\,yr$^{-1}$ there should be an upper 
limit to the grain sizes of $\sim0.1\mu$m. 
However, this assumes an isotropic distribution of material. The existence of
meteoritic titanium carbide (TiC) within presolar carbon grains necessitates
density inhomogeneities in the gas outflows \citep[c.f., inhomogenities caused 
by turbulence in hydrodynamic models of carbon-rich dust formation 
regions;][]{woitke06}.
In addition, the grains must be \lapprox 1$\mu$m to produce the 11$\mu$m 
feature, implying that 
there is a large population of small grains (but not necessarily precluding 
large grains). Constraints on grain size were also discussed by \citet{mr87}
who found  an upper limit to the grain size of $\sim$0.1$\mu$m based on 
polarization measurements of the famous carbon star, IRC+10216. On the other 
hand, \citet{Groenewegen1997} used polarization measurements to limit the 
grain size to $0.1 \leq a \leq 0.35\mu$m and suggested that very small grains 
($<$80 nm) may not exist around carbon stars (however, smaller SiC grains are found in meteorites). Meanwhile, \citet{jura94} argued 
the case for grains larger than 1$\mu$m in the circumstellar shells of 
IRC+10216.

Towards the end of the AGB the onset of the superwind may lead to mass-loss 
rates as high as a few $\times 10^{-4}$ M$_\odot$\,yr$^{-1}$ 
\citep[e.g.][]{hony2003} which could translate into larger grains 
\citep[see e.g.][and references therein]{bern05}.
However, the relationship between the evolution of carbon stars and the 
consequent evolution of grain sizes in their circumstellar shells was discussed
by \citet{Speck2005}. They argued that the increased mass-loss rates at the 
end of the AGB phase lead to smaller, rather than larger grains as suggested 
by meteoritic evidence. Even the 
highest observed mass-loss rates cannot account for formation of titanium 
carbide (TiC) grains unless the distribution of material is not 
spherically symmetric and density enhancements exist \citep{bern05}. Such a 
distribution of material makes the concept of grain size distributions even 
more complex.

The range of grain sizes used in previous radiative transfer modeling attempts 
varies hugely. 
The IRC+10216 models of \citet{griffin1990} required grain sizes limited to 
$5 \leq a \leq  50$nm. Similarly \citet{bdg95} required grains smaller than 
50nm.
\citet{Groenewegen1997} also modeled IRC+10216 and found that the spectrum was 
best fit using a single grain size of 0.16$\mu$m (rather than a distribution 
of grain sizes). The single-grain size approach was also adopted by 
\citet{Volk2000} who assume a single grain size of 0.1$\mu$m.
This is supported by the success of the early carbon star models of 
\citet{rrh83} who also used a single grain size (0.1$\mu$m).
However, since extreme carbon stars are expected to be the direct precursors  
for post-AGB objects, it may be more appropriate to consider the models of 
post-AGB stars. Such modeling efforts have found they need grains up to 
millimetre- or even centimetre- in size \citep[e.g.][]{jura2000,meixner2002}.
\citet{meixner2002} used a KMH-like distribution with 
an effective maximum grain size of 200$\mu$m (hereafter referred to as KMH200).
Meanwhile, \citet{szc97} modeled a carbon-rich post-AGB object with a 
power-law distribution of grains in the range 0.01--1$\mu$m.

Given the range of possible grain sizes and distributions, it is difficult to 
know how best to model the dust.
\citet{Groenewegen1995} argued that the specific grain-size is not important 
as long as the grains are small enough that the absorption and scattering 
properties are independent of grain size. However, given the arguments for a 
population of large grains, limiting grains to smaller than 0.1$\mu$m is 
unrealistic.

\begin{figure}[t]
%\resizebox{\hsize}{!}{%
\includegraphics[angle=270,scale=0.75]{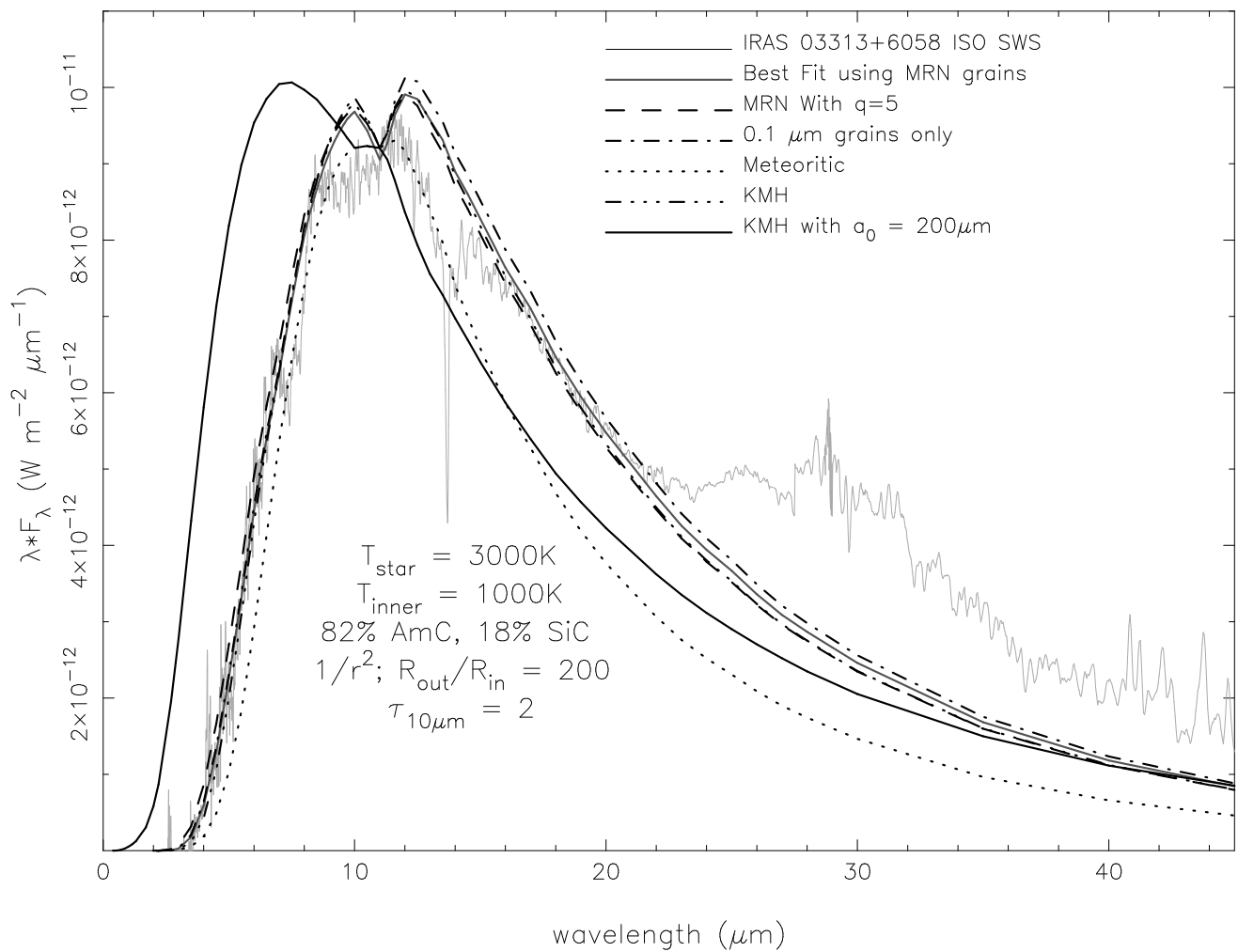}
\caption{\label{gsizefig} The effect of grain-size distributions on the model spectral energy distribution. In all cases the model parameters are identical except for the grain-size distribution. see \S~\ref{gszmodel} for detailed description of the grain-size distributions. The grey line shows the ISO-SWS spectrum of IRAS\,03313+6058 for comparison.}
\end{figure}

In order to determine the effect of the choice of grain sizes on the model 
spectra we have generated models using five additional grains size 
distributions:
1) MRN-like with a steeper power law, i.e. $a_{min} = 0.005\mu$m, 
$a_{max} = 0.25\mu$m; $q=5$ \citep[as suggested by][]{dsg89};
2) only 0.1$\mu$m-sized grains 
\cite[as used or suggested by][]{Volk2000,rrh83,Groenewegen1995,mr87};
3) the dominant meteoritic grain sizes, i.e. 0.1--1$\mu$m only;
4) the standard KMH distribution (i.e. $n(a) \propto a^{-q}e^{a/a_0}$,
$a_{min} = 0.005\mu$m, $a_{0} = 0.2\mu$m; $q=3.5$); and finally
5) KMH200: KMH with $a_0 = 200\mu$m.
The results of this study are shown in Fig.~\ref{gsizefig}, which shows the 
effect of changing the grain size distribution while keeping all other 
parameters constant.

If we examine the differences in the spectra generated by changing the 
grain-size distributions, this can be understood in terms of the absorption 
efficiency of the grains and breadth of the grain size distribution.
Changing from our default grain size distribution (MRN) to a modified MRN-like 
distribution with a steeper power law drop off ($q=5$) as suggested by 
\citet{dsg89} makes very little difference to the model spectral energy distribution (SED).
Likewise, switching from MRN to KMH has little effect on the overall shape of 
the SED.
That the MRN, $q=5$ and the KMH models are so similar is because they are 
basically weighted towards the same small grains. While the weighting is 
different, the same-sized small grains dominate the SED.
However, changing the size distribution to include larger dust grains as used 
in models of post-AGB stars \citep[e.g.][]{meixner2002} has an major effect. 
The SED shifts to peak at much shorter wavelengths. This is because of the 
reduction in the number of small grains in order to include larger grains. The 
proportion of larger 
grains is small, but the removal of the smaller grains makes it possible for 
the stellar photons to penetrate deeper into the dust shell and provides a 
large population of warmer grains, resulting in a warmer SED.
In both the KMH200 and 0.1$\mu$m cases there is a lack of very small grains 
which absorb a lot of stellar photons and change the temperature distribution 
(i.e. 
after the first layer of dust the temperature is lower, but if there are no 
small grains the stellar photons penetrate further.) Since size distributions 
similar to KMH200 are typically associated with long-lived dust disks, this 
particular distribution is not considered further.

For the meteoritic grain size distribution, the short wavelength side of the 
SED is similar to the default (MRN) model, but now the SED is much narrower. 
This can be explained by the very narrow range of grain sizes. There are no 
very small grains that can be easily heated and thus the shorter wavelength 
emission seen in KMH200 does not occur, but the lack of small grains also 
allows deeper penetration of stellar photon leading to a narrower temperature 
distribution.

This grain size study suggests that for most adopted grain-size distributions, 
the resulting SEDs will be equivalent and we assume the MRN distribution as 
``generic''. However, the meteoritic grain-size distribution can narrow the 
overall SED. 
For this reason, our modeling efforts concentrate on the generic (MRN) and ``meteoritic'' grain size distributions, where the ``meteoritic'' is taken to be an extreme amongst the range of reasonable grain-size assumptions.
The impact of choosing different grain size distributions and the implications 
of these differences for the potential errors in our models
will be discussed in \S~\ref{results}. However, essentially, most grain-size 
distributions will yield the same results except for ``meteoritic'' and 
KMH200. In both cases changes in optical depth, inner dust temperature and/or 
relative geometrical shell thickness can be manipulated to fit the spectrum.
There is a degeneracy in model fits due to the relationship between these 
three parameters that will be discussed in the next section.
As will be seen in the \S~\ref{results}, the grain size effects cannot be 
ignored.  In one source the need for ``meteoritic'' grain-size distribution 
is clear.

\subsection{Parameter space investigated}

In addition to grain sizes and radial density distribution, the variables investigated with DUSTY are the 
inner dust shell temperature (T$_{\rm inner}$), 
optical depth (specified at 10$\mu$m; $\tau_{10\mu \rm m}$), 
dust composition, and 
the geometrical thickness of the dust shell, $\xi = R_{\rm out}$/R$_{\rm in}$, 
where
R$_{\rm in}$ and R$_{\rm out}$ are the inner and outer radii of the dust 
shell, respectively. 
The optical constants for the dust components came from 
\citet{Pegourie1988,Hanner1998} and \citet{DL84} for SiC, amorphous carbon
and graphite, respectively.
In nearly all cases it was possible to generate more than one model to fit the spectra, consequently we also investigate this degeneracy in parameter space and
look for realistic ways to restrict it.

\subsubsection{Degeneracies in radiative transfer modelling}
\label{degen}

There is a clear degeneracy in the models because of the relationship between
certain parameters, e.g. optical depth and geometric shell thickness, or 
optical depth and the temperature at the inner edge of the dust shell. 
\citet{ie97} discussed these degeneracies and the relationship between the 
different input parameters in radiative transfer models, but we 
need to understand these relationship  if we are to 
understand what our models mean.

The wavelength-dependent optical depth, $\tau_\lambda$ is defined by:

\begin{equation}
\label{tau0}
d\tau_\lambda = \rho(r) \kappa_\lambda dr 
\end{equation}

\noindent
where $\rho(r)$ is the density of the absorbing/scattering particles 
(i.e. the  dust grains) and $\kappa_\lambda$ is the wavelength dependent 
opacity of the assemblage of particles along the line of sight.

\[ \tau_\lambda = \int^{R_{\rm out}}_{R_{\rm in}}{\rho(r) \kappa_\lambda dr} \]

\noindent
where $R_{\rm in}$ is the inner dust shell radius and $R_{\rm out}$ is the 
outer dust shell radius.
While $\kappa_\lambda$ is dependent on the density distribution of the grains,
maintaining the size, shape and composition 
(and crystal structure) of the grains means that $\kappa_\lambda$ will not 
change significantly. 
For simplicity we assume   $\kappa_\lambda$ remains constant.
In our models we assume the density of the dust shell drops off as 1/$r^2$ 
from the central star. In addition, our models specify the relative geometrical
thickness of the dust shell as $\xi = R_{\rm out}/R_{\rm in}$. Thus we get:

\begin{equation}
\label{tau1}
\tau_\lambda = \kappa_\lambda \left[ \frac{\xi - 1}{\xi R_{\rm in}} \right]
\end{equation}

\noindent
The value of $R_{\rm in}$ is set by the values we input for the star's 
effective temperature and the inner dust radius (or condensation
 temperature, $T_{\rm inner}$).
If we know the temperature and luminosity of the star, we can use a 
$T(r) \propto r^{-\frac{1}{2}}$ temperature distribution to determine the
relationship between  $T_{\rm inner}$ and $R_{\rm in}$ and 
substituting into Eq.~\ref{tau1} we get:

\begin{equation}
\label{tauTxi}
\tau_\lambda = \sqrt{\frac{16 \pi \sigma}{L_\star}}~\kappa_\lambda T_{\rm inner}^2 \left[ \frac{\xi - 1}{\xi}  \right]
\end{equation}

Here we assume that the grains are blackbodies because they are largely 
carbon. Including the albedo would allow for a more accurate calculation, but 
this would depend on detailed dust parameters (like crystal structure), and the error incurred by our assumption is small 
(i.e.\ significantly less than an order of magnitude).

Therefore, according the Eq.~\ref{tauTxi}, for a dust shell with constant 
relative shell thickness ($\xi$) the optical depth 
should increase with the square of the inner dust temperature. Alternatively, 
if the inner dust temperature is fixed, then increasing the relative shell 
thickness should decrease optical depth  a little (as $(\xi - 1) / \xi $). 
This latter effect becomes negligible for large geometric sizes.

\subsection{Determining the Dust Condensation Temperature}
\label{constraints}

\subsubsection{Theoretical dust condensation models}
\label{condmod}

As seen in the previous section, the value of the dust condensation 
temperature is a source of degeneracy on radiative transfer modeling.
In order to reduce this degeneracy we turn to dust condensation 
theory to determine a theoretical dust condensation temperature 
(i.e. $T_{\rm inner}$) that is appropriate for our stars.

As discussed in \S~\ref{meteor}, many presolar grains can be attributed to 
carbon stars and are valuable resources for investigation of dust formation 
regions. For example, in presolar grains titanium carbide (TiC) is found in 
the center of carbon (C) 
grains from AGB stars, but only one SiC grain has been found coated in carbon 
\citep{cn04, bern05}.
Consequently there have been many studies of the theoretical condensation 
sequence in Galactic carbon stars in attempts to constrain the physical 
parameters of the dust condensation regions.
These studies showed how the condensation sequence of C, TiC and SiC is 
dependent on various parameters, most notably C/O ratio and gas 
pressure\footnote{%
Gas pressure is a measure of the mass-loss rate ($\dot{M}$) convolved with the 
photospheric temperature (T$_\star$) and outflow velocity ($v_{\rm exp}$).} 
\citep{lf95,sw95}. 
%
%\citet{sw95} found that graphite (carbon) condensation 
%temperature (T$_{cond}$) is strongly dependent on C/O, while the T$_{cond}$ 
%for TiC and SiC are not.
%Conversely, the condensation temperature of carbon is more or less 
%independent of gas pressure, but T$_{cond}$ for TiC and SiC varies strongly 
%with gas pressure. However, \citet{lf95} found that C, TiC and SiC are all 
%strongly dependent on both gas pressure and C/O ratio.
%
 \citet{sw95} argued 
that if carbon forms at a higher condensation temperature, closer to the star 
than SiC, there is a significant decrease in 
the amount of carbon available in the gas, and thus SiC and C do 
not form simultaneously, resulting in naked SiC grains. Therefore, for Galactic
sources, the condensation sequence in the majority of carbon stars should be 
TiC --- C --- SiC, in order to produce the coated TiC grains and uncoated 
(naked) SiC grains seen in the meteoritic presolar grains samples. 
Observational evidence for naked SiC grains is discussed by \citet{Speck2005}.
\citet{sw95} argued that from kinetic and stellar model considerations, dust 
grains should form in the pressure range 2 $\times 10^{-7} <$ P $< 
4 \times 10^{-5}$ bars\footnote{%
The expected range of gas pressures in the dust formation zone for 
O-rich AGB stars in the LMC is 10$^{-7}$ bars \lapprox P \lapprox 10$^{-4}$ 
bars \citep{dijk05}}.
%
%Therefore, to form dust in the sequence TiC --- C --- SiC, they need 
%1.04 $<$ C/O $<$ 1.2. For Galactic carbon stars

\citet{lf95} also modeled the effect of C/O and pressure on the condensation 
sequence in carbon stars, as well as the effect of {\it s}-process and 
nitrogen  abundances. They also briefly discuss the effect of metallicity.
The general trends in condensation temperatures are: 
(1) all condensation temperatures decrease as the gas pressure decreases; and 
(2) At C/O $>$ 1 the condensation temperature of graphite increases with C/O 
(for a given pressure and otherwise constant composition).

Figure~\ref{lfptspace} shows how the condensation temperature of carbon and 
SiC vary with the gas pressure in the dust condensation 
zone. For the range plotted carbon always forms first from a cooling gas. 
Very high pressures are required for SiC to form before carbon.
For solar metallicity and C/O = 1.05, SiC forms before carbon for 
P $\geq 3.4 \times 10^{-5}$ bars. 
As C/O increases, the minimum pressure required to form SiC first increases.
Above C/O $\sim$1.5 carbon always forms before SiC. The exact C/O ratio at 
which 
the carbon forms before SiC depends on pressure. Therefore, in order to 
account for observations of SiC features in the Galaxy, and the presolar 
grain record, we can restrict the P--C/O space such that, for low C/O the gas 
pressure must remain low, but for higher C/O the pressure can be higher. 
This can be used to constrain the dust forming environment around Galactic 
carbon stars. 

Studies of carbon star spectra in the Galaxy and the Magellanic Clouds have 
lead to different interpretations with respect to the condensation sequence.
\citet{Lagadec2007} argued for a sequence in which SiC forms before C in the 
Galaxy, whereas SiC and C form together in the Large Magellanic Cloud (LMC) 
and the sequence is reversed (C, then SiC) in the Small Magellanic Cloud (SMC). However, they also suggest that the change in the strength of the features is 
due to the lower number of Si atoms available for SiC formation.
The proposed Galactic condensation sequence is at odds with both the models 
and the meteoritic evidence. \citet{leisenring} support the condensation 
sequence in which C forms before SiC for the 
Galactic carbon stars, while finding that the Magellanic Clouds tended to form 
SiC first, followed by simultaneous condensation of SiC and C.
\citet{Speck2006} used the \citet{lf95} model to explain an unusual LMC carbon 
star spectrum, which suggests that the condensation sequence is sensitive to 
both metallicity and mass-loss rate.

\subsubsection{P-T space in the condensation zone around extreme carbon stars}
\label{ptcalc}

In order to constrain the input parameters to our model we need to be able to 
determine the pressure-temperature space in the dust condensation zone.
For a mass-losing star with a mass-loss rate $\dot{M}$ and 
an expansion velocity of $v_{\rm exp}$, 
the density $\rho$ of the circumstellar shell 
at a radius $r$ is given by:

\[ \rho = \frac{\dot{M}}{4 \pi r^{2} v_{\rm exp}} \]

If we know the temperature and luminosity of the star and the composition of 
the outflowing material we can combine this information with the Ideal Gas Law 
and a $T(r) \propto 1/\sqrt{r}$ temperature distribution to determine 
the
gas pressure at the condensation radius, where the condensation radius is the
distance from the star where the gas has the condensation temperature 
($T_{\rm cond}$).

For simplicity, the solid and gas phases are assume to be at the same 
temperature. While this is clearly a simplification \citep[e.g.][]{chigai}, 
the temperature difference is small compared to the difference needed to 
significantly affect dust formation.
We assume that  most of the outflowing material is atomic hydrogen. In fact it will probably be a mixture of atomic and molecular hydrogen (H$_2$) since H$_2$
forms around 2000\,K and the temperature in the outflow is decreasing from 
the stellar surface temperature of $\sim$3000\,K to the dust condensation 
temperature in the 1000--1800\,K range. 
%
%However, without dust grains the gas may not form H$_2$ efficiently, so we assume for simplicity that the all the gas is atomic hydrogen. 
An entirely molecular hydrogen gas would halve 
the gas pressure compared to the the atomic gas. 
However, we also assume published CO outflow velocities, which reflect the 
speed of the outflowing material after radiation pressure acceleration. 
Adopting the pre-dust-formation outflow speed (\lapprox 5km/s), would increase 
the pressure.
Thus using published values for our sample stars' luminosities, 
mass-loss rates and expansion velocities, we can estimate where their dust 
condensation zones fall in P--T space.

\subsubsection{Comparison of P--T space for dust condensation models and sample 
stars}
\label{comppt}

\begin{figure}[t]
%\resizebox{\hsize}{!}{%
%
~~~~~~~\,\includegraphics[angle=270,scale=0.6]{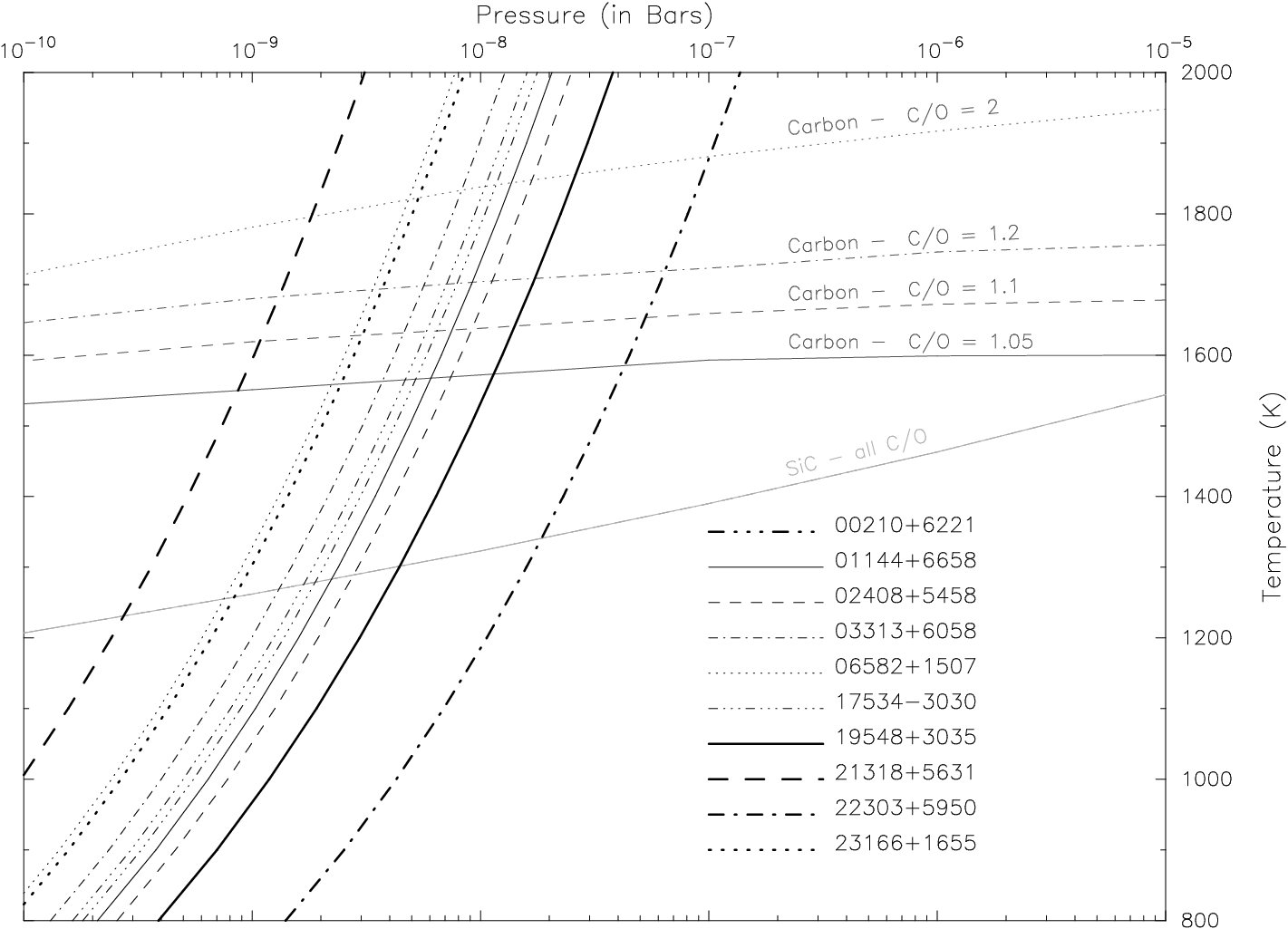}\\
%\vspace{-25mm}
%\hspace{25mm}
\includegraphics[angle=270,scale=0.6]{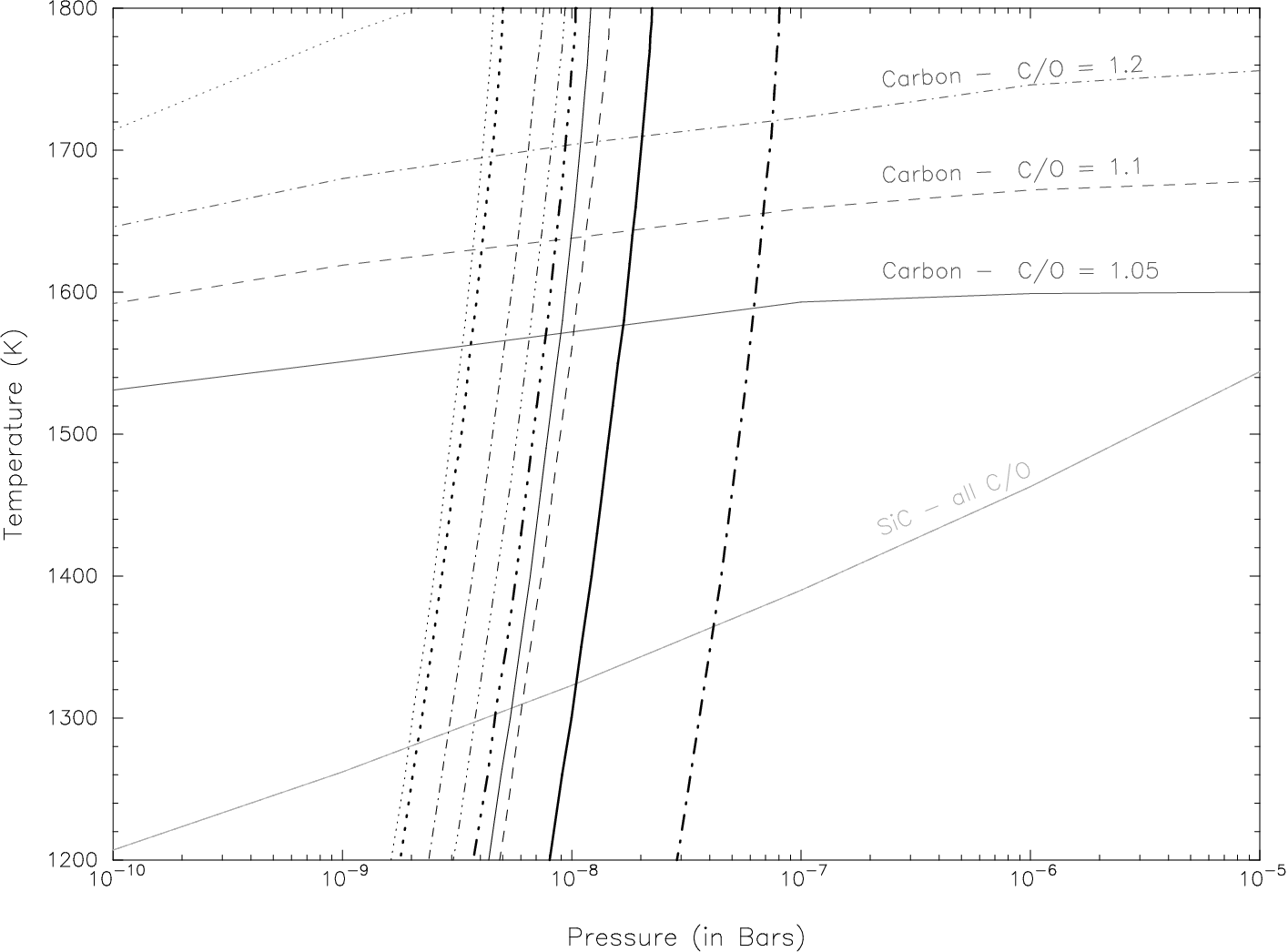}
\caption{\label{lfptspace} Pressure-Temperature space for the dust condensation region around extreme carbon stars (assumes solar metallicity). Grey lines indicate the condensation temperature for a given pressure as calculated by \citet{lf95}. Black lines indicate the P--T paths for the otuflowing gas from our sample stars; ({\it top}) as calculated from the published CO mass-loss rates and expansion velocities (see \S~\ref{ptcalc} for details); and ({\it bottom}) as calculated from the modeled dust temperature radial profile. }

%The best model inner dust temperature is indicated by a triangle on each P--T track.}
\end{figure}

Figure~\ref{lfptspace} shows how the P--T space for the dust condensation zone 
for our target stars compares to condensation models in P--T space.
It is clear the pressure is never high enough for SiC to form before 
carbon.  Since our sample stars are expected to be the carbon stars with the 
highest mass-loss rates, it implies that this NEVER happens in Galactic carbon 
stars. 
This agrees with the meteoritic evidence which suggests that SiC does
not get coated in carbon and supported the idea that variations in the 
strength of the SiC feature are related to self-absorption.

The dust condensation temperature is dependent on the pressure in the gas from 
which the dust forms.
In addition, the P--T space occupied by the target sources suggests 
that carbon grains will form at temperatures \gapprox 1600\,K. 
If the C/O is 
very high, then graphite could form as high as $\sim$1800\,K. 
However, Galactic 
carbon stars for which the C/O ratios have been measured show that it is 
in the range 1 to 1.8, with a mean C/O ratio of $\sim$1.15 
\citep{lamb86, olof93a, olof93b}. The precise C/O for our sample is not known,
but even low C/O stars yield $T_{\rm cond}$ \gapprox 1550\,K 
(see Fig.~\ref{lfptspace}).
Previous models of carbon stars have assumed much lower inner dust 
temperatures.
The pressures and temperatures in the gas around these 
stars meet the criteria for forming carbon dust at $T_{\rm cond}$ \gapprox 
1600\,K, (precise temperature depends on the C/O ratio). Therefore it should 
form at these high temperatures.

Once dust starts to form, the radial temperature profile will change due to 
absorption of starlight by dust grains.
Therefore, we use the comparison above only to determine the inner dust radius.
Using radial temperature profiles from our models we show that the temperature 
drops significantly more rapidly than $1/\sqrt{r}$ (see Fig.~\ref{lfptspace}), 
which suggests that the 
next condensate (SiC) forms fairly close to the inner dust radius and thus 
mitigates the problem with DUSTY that the grains are assumed to be co-spatial. 
Indeed the models show that the temperature in the dust shell drops to the SiC 
formation temperature at $\sim$1.3R$_{\rm in}$ for all cases, which is small 
compared to the shell thickness, even for the thinnest shells.

\subsection{Constraining Dust Composition}

Since our stars are carbon-rich, we limit the models to only including 
carbonaceous species such as graphite, amorphous carbon and silicon carbide.
The choice of carbon grains is equivocal. We do not know whether the grains 
composed mostly of carbon are glassy, poly-nanocrystalline or well-ordered.
The meteoritic presolar ``graphite'' grains suggest that circumstellar dust 
can contain either well-ordered graphite or poly-nanocrystalline-graphene 
grains. It should be noted, however, that even the disordered graphene sheets 
are considered to be more graphitic than amorphous. 

Amorphous carbon essentially consists of a mixture of $sp^2$ (graphite-like) 
and $sp^3$ (diamond-like) carbon bonds. On heating, $sp^3$ bonds tend to 
convert to $sp^2$ bonds, thus graphitizing the amorphous carbon, but it will 
remain dense like the amorphous phase 
\citep[2.8\,g/cm$^3$][]{comelli,saada,kelires}. 
At $\sim$1300\,K the fraction of 
graphitic bonds is $\sim$90\%. 
Comparison of the gas pressure in the circumstellar outflows to graphite 
formation temperature in \S~\ref{comppt} shows that graphite should be able to 
form at temperatures significantly above 1300\,K. Even if solid state carbon 
forms as a chaotic solid, at these temperatures it will quickly anneal to a 
graphitic form.
Consequently, we argue that graphite is the better choice of carbonaceous 
material for modeling extreme carbon star dust shells.

The limitations of the use of the \citet{Pegourie1988} data are discussed in 
\citet{Pitman2007}. Clearly, this data cannot produce the broad absorption 
feature seen in the observations. However, we can use the relative changes in 
composition from star to star as a guide to understanding why these stars 
have the features we see.

\section{Radiative Transfer Modeling Results}
\label{results}

\begin{figure}[t]
%\resizebox{\hsize}{!}{%
\includegraphics[angle=270,scale=0.7]{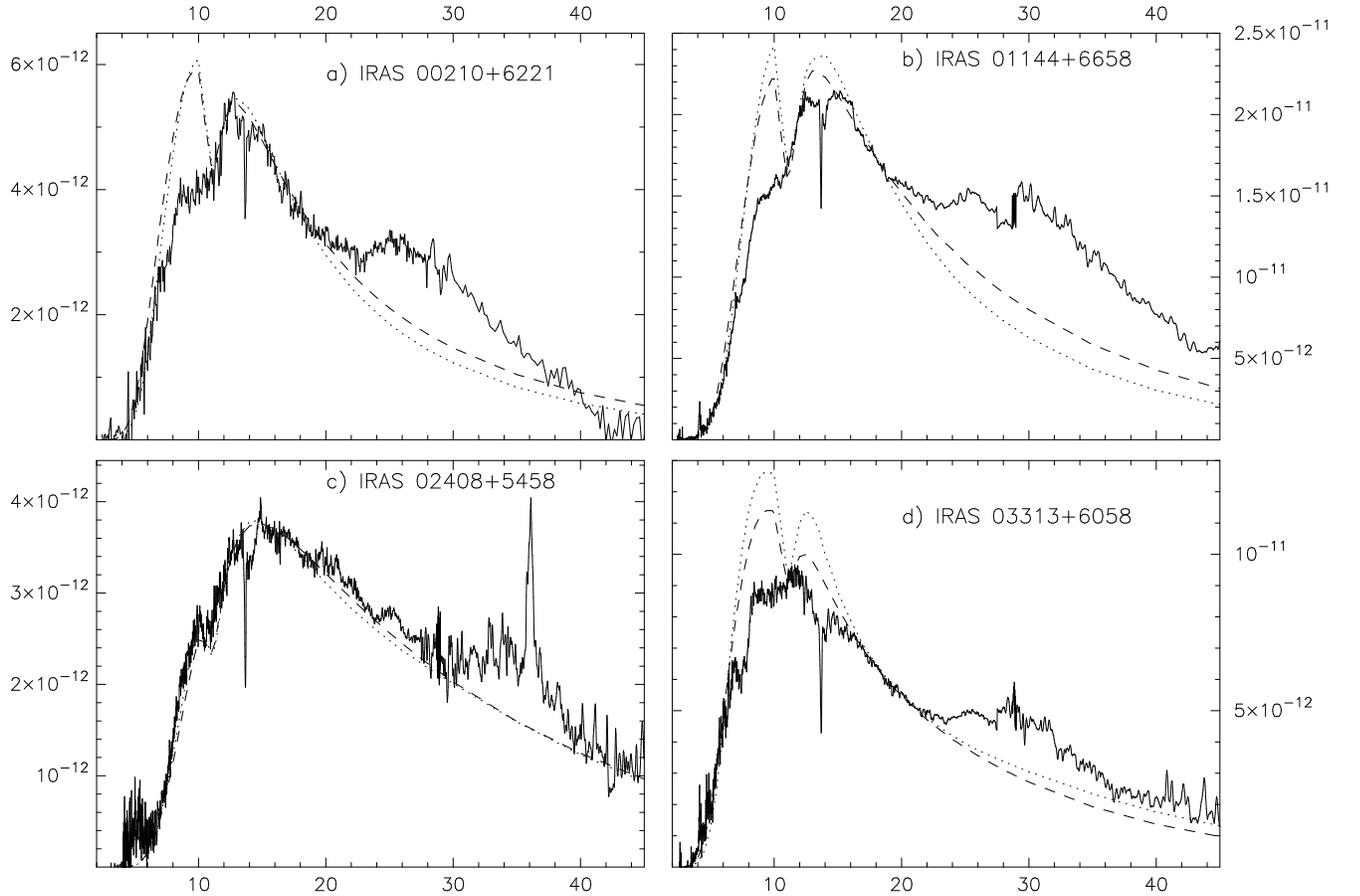}
\caption{\label{modelfit1} Best fit models (part 1).
solid line = ISO-SWS spectrum;
dashed line = best fit model using MRN grain-size distribution;
dotted line =  best fit model using ``meteoritic'' grain-size distribution;
$X$-axis is wavelength ($\mu$m);
$y$-axis is flux ($\lambda F_\lambda$) in W\,m$^{-2}$.
In all cases, $T_\star$=3000\,K and $T_{\rm inner}$=1800\,K. 
Model parameters are listed in Table~\ref{tabfit1}}
\end{figure}

\begin{figure}[t]
%\addtocounter{figure}{-1}
%\resizebox{\hsize}{!}{%
\includegraphics[angle=270,scale=0.7]{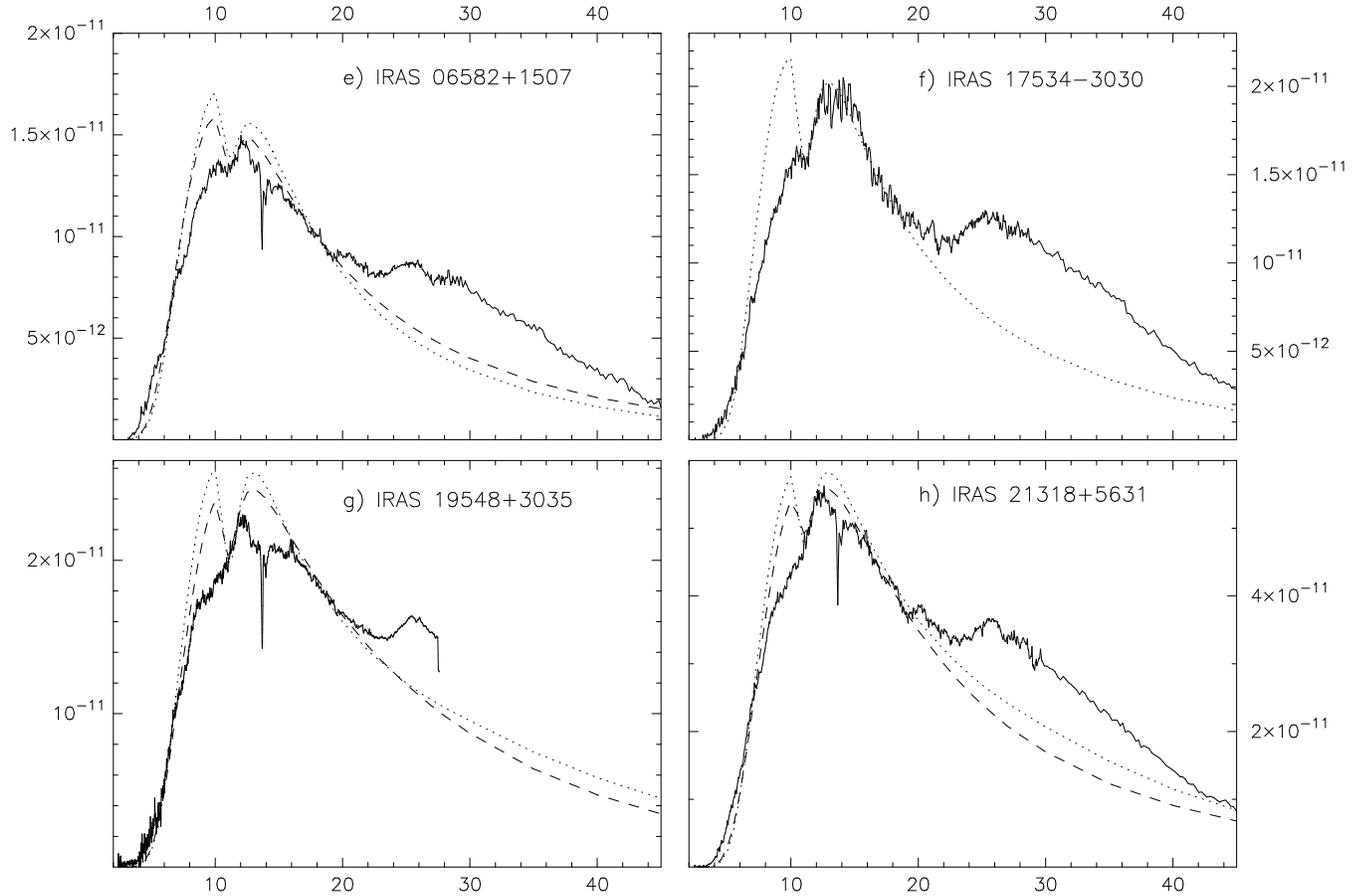}
\caption{\label{modelfit2} Best fit models (part 2) 
solid line = ISO-SWS spectrum;
dashed line = best fit model
$x$-axis is wavelength ($\mu$m);
$y$-axis is flux ($\lambda F_\lambda$) in W\,m$^{-2}$.
In all cases, $T_\star$=3000\,K and $T_{\rm inner}$=1800\,K. 
Model parameters are listed in Table~\ref{tabfit1}}
\end{figure}

\begin{figure}[t]
%\addtocounter{figure}{-1}
%\resizebox{\hsize}{!}{%
\includegraphics[angle=270,scale=0.7]{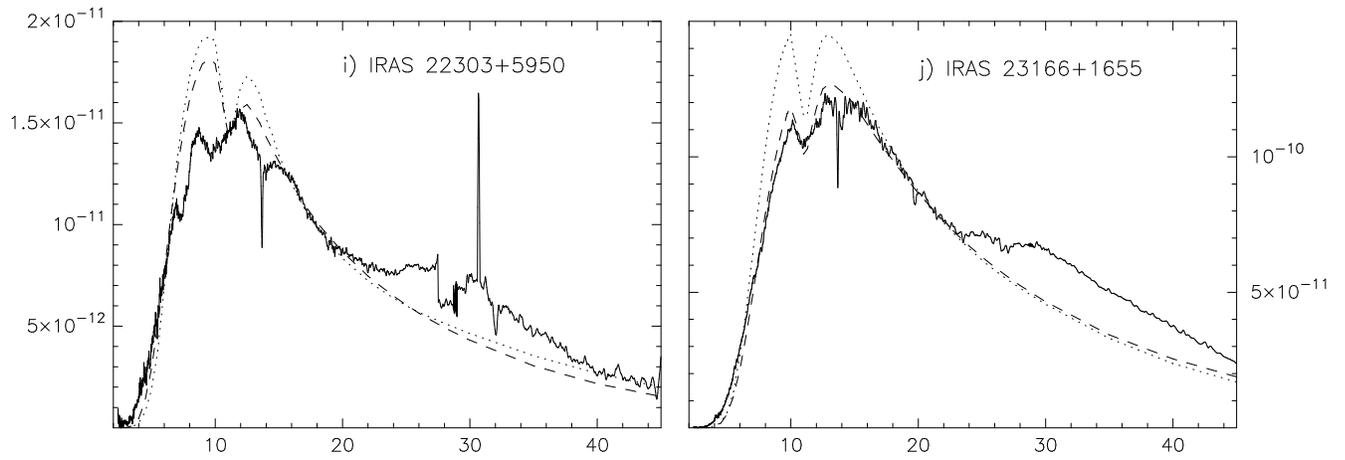}
\caption{\label{modelfit3} Best fit models (part 3) 
solid line = ISO-SWS spectrum;
dashed line = best fit model
$x$-axis is wavelength ($\mu$m);
$y$-axis is flux ($\lambda F_\lambda$) in W\,m$^{-2}$.
In all cases, $T_\star$=3000\,K and $T_{\rm inner}$=1800\,K. 
other parameters are indicated in the legend. 
Parameters of these models are also listed in Table~\ref{tabfit1}}
\end{figure}

The results of the radiative transfer modeling can be seen in 
Fig.~\ref{modelfit1}, \ref{modelfit2} and \ref{modelfit3}. 
The parameters used in each case can be seen in Table~\ref{tabfit1}.
For each source, we have produced models using both MRN (generic) and 
``meteoritic'' (extreme) grain size distributions.

\subsection{The effect of grain-size distribution}
\label{gsizemod}

Table~\ref{tabfit1} shows the model input parameters for both grainsize distributions. 
In all cases switching from MRN to ``meteoritic'' grainsizes leads to the need for increased SiC component 
(compared to graphite), typically requiring a three- or fourfold increase in the SiC fraction.

There are other general trends including the need for decreased optical depths ($\tau_{10\mu \rm m}$) and 
increased geometric shell thickness. However, these trends do not hold for all objects. 
In the cases of IRAS\,00210+6221 and IRAS\,06582+1507, the only difference in parameters between 
the different grain size models is the fraction of SiC. 
IRAS\,01144+6658 is the only source for which optical depth was increased and shell thickness was 
decreased in the ``meteoritic'' model.

\subsection{Dust Shell Thicknesses}
\label{timescalesect}

\begin{figure}[t]
%\resizebox{\hsize}{!}{%
\includegraphics[angle=270,scale=0.7]{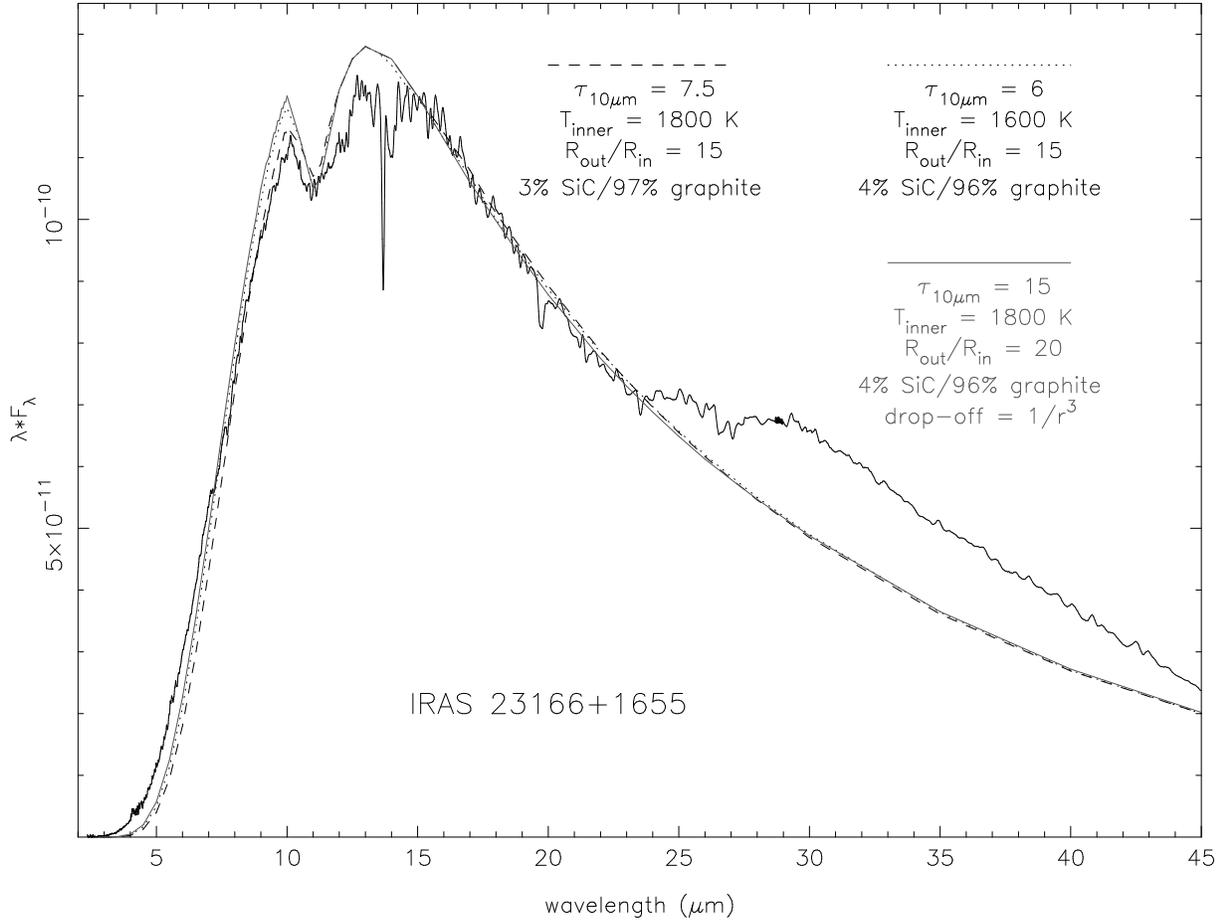}
\caption{\label{modelfit23166} Degeneracy in the best fit models for 
IRAS\,23166+1655.  
solid line = ISO-SWS spectrum;
dashed line = best fit model with $T_{\rm inner}$=1800\,K
dotted line = best fit model with $T_{\rm inner}$=1600\,K
$X$-axis is wavelength ($\mu$m);
$y$-axis is flux ($\lambda F_\lambda$) in W\,m$^{-2}$.
In both cases, $T_\star$=3000\,K 
other parameters are indicated in the legend. }
\end{figure}

As discussed in \S~\ref{degen}, there is a degeneracy between relative shell 
thickness, inner dust temperature and optical depth. The inner dust 
temperature variability has been limited to 1550--1800\,K by 
theoretical considerations (see \S~\ref{condmod}). 
It is possible to tweak parameters such as 
shell thickness, inner dust temperature and optical depth and get almost 
identical models. Fig.~\ref{modelfit23166} shows two almost identical model 
spectra for IRAS\,23166+1655  with different inner dust temperatures, 
optical depths and shell 
thicknesses. However, reducing the inner dust temperature merely requires 
reduction of the optical depth.
While increasingly geometrically large dust shells can be accommodated by 
decreasing optical depth (see Eq.~\ref{tauTxi}), at some point, this also 
breaks down, as it leads to a significant population of colder grains which 
emit too much at long wavelengths. In this way we can use the models to place 
to an {\it upper limit to the shell thickness}. 
Table~\ref{obsparam2} lists the published expansion velocities for our 
extreme carbon stars.
From the model parameters, the physical value for 
$R_{\rm out}$ can be calculated,
which gives the physical size of the dust shell. 
Since the models give an upper limit to the shell thickness, 
this upper limit to $R_{\rm out}$ together with the expansion velocity ($v_{\rm exp}$) 
was used to calculate the 
time since the outermost edge of the shell was ejected from the star.
The resulting ages of the dust shells are listed in Table~\ref{timescaletab}.

As can be seen in Fig.~\ref{gsizefig}, it is possible to accommodate thick 
shells, if we assume a much lower $T_{\rm inner}$ (1000\,K), and assume that 
the dust grain are composed of amorphous rather than graphitic carbon, but 
these parameter values are precluded by the theoretical constraints
on $T_{\rm inner}$ and composition/crystal structure discussed above.
Clearly, since our constraints are theoretical, they may change as hypotheses 
are refined.

The timescales for increased mass-loss are model dependent. As can be seen in 
Table~\ref{tabfit1} using a ``meteoritic'' grain size distribution generally 
requires a geometrically thicker shell, as well as lower values for 
$\tau_{10\mu \rm m}$ and higher percentages of SiC. However,
timescales associated with our dust shells are always very short (less than a 
few thousand years) regardless of grain-size distribution. 
Moreover, our timescales are consistent with those derived for the 
``superwind'' 
seen in post-AGB objects \citep[e.g.][]{skinner97,meixner2004}.
These timescales are too short to be associated with the theoretical
superwind \citep[e.g.][]{vw93}, 
which is expected to last up to $\sim$10\% of the duration of the 
thermally-pulsing AGB phase.
However, if we compare the number of extreme carbon stars to the 
total number of carbon stars in the Galaxy, we find that extreme 
carbon stars constitute only $\sim$0.1\% of the total 
C-star population. The thermally pulsing AGB phase is expected to last
$\sim 10^6$ years, and the time a star spends as a C-stars is even 
shorter \citep[e.g.][]{Lagadec2007}. Consequently, we might expect the 
dust-obscured phase to only last 
$\sim 10^3$ years. This is consistent with the model timescales, 
although the MRN-models still appear to have very short timescales.
\citet{LZ08} suggested that the trigger for the superwind is a combination of 
luminosity and carbon abundance. Although the duration of the C-star phase 
maybe be $\sim 3 \times 10^5$\,years, for much of this time, a C-star will be
below the critical carbon abundance required to drive the superwind and 
obscure the star. 

How does the extreme carbon star phase fit into the broader C-star evolution?
Many, if not most, carbon-rich post-AGB stars showed marked axisymmetric 
morphologies \citep{meixner1999,waelkens,sahai,soker}. 
The extreme carbon stars are expected to be the 
direct precursors of these objects \citep[e.g.][]{skinner98}, but presently 
show little evidence for 
axisymmetry\footnote{one would expect to see more near-IR emission in the 
spectrum of strongly axisymmetric objects due to scattering of starlight 
escaping in the bipolar axis direction.}.
The cause and timing of this axisymmetric structure is not known, but is 
believed to occur at the very end of the AGB phase.
\citet{ds06} showed that significant axisymmetry is not expected to develop 
until the last few tens or hundreds of years of the superwind phase.
The onset of axisymmetry also leads to an optically thicker toroid of dust, as 
the circumstellar shell becomes equatorially enhanced.
We suggest that the very short timescales associated with our model results 
may indicate that the extreme AGB stars are in the process of developing 
axisymmetry, but that this has not developed to the point of allowing large 
amounts of NIR scattered light into the spectrum.

\subsection{Dust Shell Density Distribution}

In \S~\ref{rtmodeling} we argued the case for maintaining a dust density 
distribution that follows a 1/$r^2$ law. However, it has been argued that the
increasing stellar luminosity and the onset of the superwind phase should 
give rise to a steeper density drop-off.

Following the same arguments as shown in \S~\ref{degen}, 
giving rise to Eq.~\ref{tauTxi} we can derive an equation in which the 
exponent of the density power law is a variable.
This gives us the relationship between the optical depth ($\tau_\lambda$),
inner dust temperature ($T_{\rm inner}$), relative shell thicks ($\xi$) and 
the exponent of the density power law ($x$):
\begin{equation}
\label{tau_rTxi}
\tau_\lambda = \kappa_\lambda \left[ \frac{\xi^{x-1}-1}{(x-1)\xi^{x-1}} \right]\left(\sqrt{\frac{16 \pi \sigma}{L_{\star}}} T_{\rm inner}^2 \right)^{x-1}   
\end{equation}

\noindent
where $\kappa_\lambda$ is the wavelength dependent opacity of the assemblage 
of particles along the line of sight;
$L_\star$ is the luminosity of the star; and
$\sigma$ is the Stefan-Boltzmann constant.

As an example of the effect this, we assume the exponent, $x$ = 3 
(which is the highest value in the previous models).
In this case, the relative shell thickness dependence dwindles, and the 
optical depth is essential strongly dependent on the inner dust 
temperature only. 
Since the previous models have severely underestimated the inner 
dust temperature, they have also underestimated the optical depth
\citep[which may explain][]{Groenewegen1995}.

The lack of dependence on shell thickness at $x > 2$ leads to a situation 
where most of the dust is essentially confined to a region close to the star 
and the outer dust becomes negligible, and so this confinement to the inner 
region is effectively the same as our assumption that there was a 
sudden increase in mass loss in the recent past.
However, we attempted to model our sources with a $1/r^3$ radial 
density distribution and found that we cannot match the shape of the SED 
without restricting the geometrical shell thickness.
Using the $1/r^3$ radial density distribution with the MRN grain-size
distribution and $R_{\rm out}/R_{\rm in}$ \gapprox 20 produces a SED that has 
too much emission longwards of $\sim20\mu$m. 
Therefore, even if the radial density distribution is indicative of a steep 
increase in mass-loss rate the shells still need to be geometrically thin, and 
the shell thickness and percentage of SiC needed are essentially the same as 
for the  $1/r^2$ models
(see Fig.~\ref{modelfit23166} for an example.) 
If the ``meteoritic'' grain-size distribution is used in conjunction with a 
$1/r^3$ drop-off, the observed SED cannot be matched. The model SED 
becomes too narrow, and increasing the outer dust-shell radius does not help.

\subsection{IRAS~17534$-$3030}
\label{17534}

IRAS\,17534$-$3030 is exceptional amongst the present sample of extreme carbon stars, 
and certainly requires more discussion. 
As can be seen in the flux-calibrated and continuum-divided spectra in Fig.~\ref{obsfig}, 
IRAS\,17534$-$3030 does not exhibit the usual molecular absorption bands around 13.7$\mu$m 
(due the C$_2$H$_2$) and shortwards of $\sim8\mu$m (due to both C$_2$H$_2$ and HCN), 
which are present in the other sample sources. Its 11--13$\mu$m feature is intermediate 
between the narrow feature exemplified by IRAS\,23166+1655 and the broad feature 
exemplified by IRAS\,19548+3035, indicating that at least some of whatever substance 
causes the broadening is present around IRAS\,17534$-$3030.
One of the suggested carriers for the short wavelength broadening of the 10-13$\mu$m 
absorption feature is C$_3$ \citep{zijl06}. However, C$_3$ is expected to be photospheric, 
rather than circumstellar, which probably precludes its detection in optically obscured stars.
Moreover, the theoretical spectrum of C$_3$ 
from \citet{jorg} shows a strong absorption close to the $\sim5\mu$m CO 
line, which is stronger than the $\sim11\mu$m feature. The spectrum of IRAS\,17534$-$3030
does not show this 5$\mu$m absorption band. This, together with the lack of other 
molecular absorption feature provides evidence that the
broadening of the 10-13$\mu$m feature is not molecular in origin.

In addition to the lack of molecular absorption in its spectrum, 
IRAS\,17534$-$3030 is unique is another way: it cannot be modeled with the MRN grain-size distribution.
Modeling of this source requires the ``meteoritic'' grain-size distribution because of its narrow SED.
Whereas our other sources can be fitted with either MRN or ``meteoritic'' grains, IRAS\,17534$-$3030 
cannot.

\subsection{Impact of the dust condensation temperature}

The increase in the inner dust temperature from \lapprox\,1000\,K in previous 
models up to 
\gapprox\,1600\,K decreases the inner dust radius significantly.
Consequently the flux of energy from the star hitting the inner dust radius 
is increased, leading to a greater acceleration and consequently a more 
effectively dust-driven wind. If graphite is formed by annealing of amorphous 
carbon, the grain density should remain high (with $\rho$=2.8g/cm$^3$) and 
the increased acceleration is entirely due to increases flux of stellar 
photons.
However, if graphite forms directly, rather than by 
annealing of amorphous carbon, the radiation pressure effect is further 
enhanced, because graphite grains generally have a 
lower density than amorphous carbon. Assuming 0.1$\mu$m-sized grains,
the acceleration felt by graphite grains (with $\rho$=2.2g/cm$^3$; c.f. 
meteoritic presolar grains; see \S~\ref{presolc}) at 
T$_{\rm inner}$ = 1800\,K is 13 times greater than the 
acceleration felt by amorphous carbon grains (with $\rho$=2.8g/cm$^3$) 
at T$_{\rm inner}$ = 1000\,K.

\subsection{Correlations between observed and model parameters}
\label{correlsect}

In the course of this investigation we have compiled a large number of 
parameters for these stars. For instance, Table~\ref{obsparam} lists the peak 
position, peak strength, FWHM and equivalent width 
of the $\sim11\mu$m absorption feature.
Table~\ref{obsparam2} lists the published mass-loss rates, luminosities and 
expansion velocities for our sample stars.

In order to understand the physical conditions that give rise to the observed 
11$\mu$m absorption feature, we have sought correlations between various 
parameters associated with the sample stars (as found in 
Table~\ref{obsparam}, \ref{obsparam2}, \ref{tabfit1}).
We have looked for correlations between each of the following:
mass-loss rate (from CO observations); 
expansion velocity (from CO observations);
stellar luminosity;
strength of the observed absorption;
FWHM of the absorption;
barycentric position of the absorption;
equivalent width of the absorption;
and model parameters.

%Loup et al 1993 
\citet{Loup1993} showed that, for low mass loss rates there is a 
simple relationship between the [25]-[12] color and mass-loss rates. 
This breaks down at mass-loss rates \gapprox $10^{-5}$\,M$_\odot$yr$^{-1}$. 
It has been suggested that such high mass-loss rate stars at CO-emission 
deficient due to either saturation effects, low kinetic temperatures or 
possibly dramatic recent increases in mass-loss. If we extrapolate the 
trends from low mass-loss rates to determine mass-loss rates from the 
[25]-[12] color we find that these objects should have a mass-loss rate 
in excess of $10^{-4}$\,M$_\odot$yr$^{-1}$, consistent with the high 
modeled optical depths. Because of this relationship, we sought 
correlations between the various observed and modeled parameters and 
the [25]-[12] and [60]-[25] colors. This search yielded only one correlation: 
between the [60]-[25] colors and the dust mass-loss rate from previous models.
It is possible that a better correlation may be found using the 
{\em Manchester Method} i.e. the [6]-[9] color 
\citep[e.g.][]{Sloan2006,zijl06}, however, we suspect that the lack of 
correlation arises because of the intrinsic degeneracy in the modeling.

In \S~\ref{sicabs} and \S~\ref{obssect} 
we discussed the possibility that the broadening of the 
10--13$\mu$m feature might be due to silicate dust.
With this in mind, we re-measured the position  and strength of the 11$\mu$m 
feature assuming that the short wavelength wing is due to silicate. 
This involved
measuring the feature-to-continuum strength at 9.7 and 11.3$\mu$m. These data
are tabulated in Table~\ref{obsparam} and were also included in the 
investigation of correlations between parameters.

\begin{figure}[t]
%\resizebox{\hsize}{!}{%
\includegraphics[angle=270,scale=1.0]{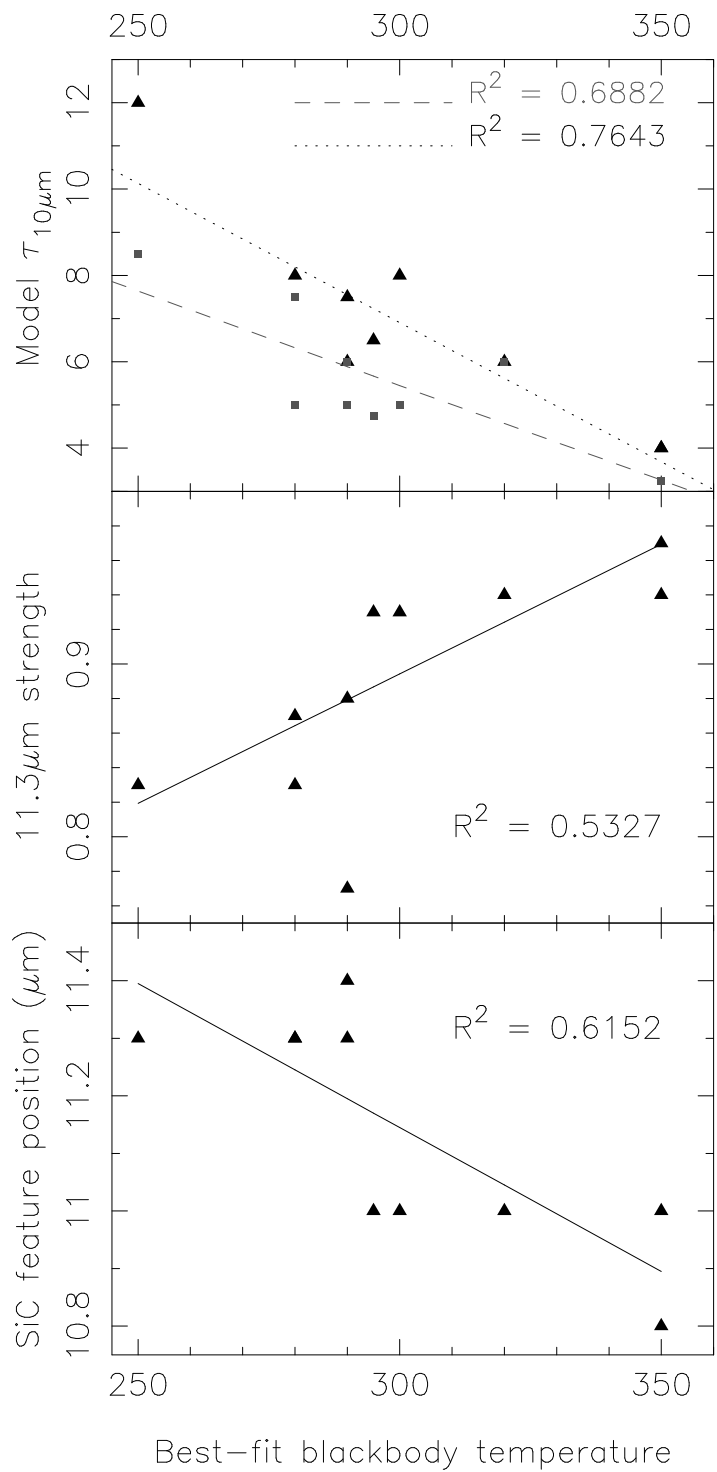}
\caption{\label{correlfig} Correlations between the best-fit 
blackbody temperature and 
({\it top}) modeled optical depth;
({\it middle}) 11.3$\mu$m feature-to-continuum ratio; and
({\it bottom}) SiC feature position.
Solid lines represent the linear regression fit through the points.
In the top figure, the dotted line represents the linear regression for model fits with MRN grainsize distributions.
Grey points are for meteoritic grainsize models, and grey dashed line is the linear regression fit through the points.}
\end{figure}

One pair of parameters that yielded a correlation were
the best-fit blackbody temperature and the model optical depth,
which is demonstrated in Fig~\ref{correlfig}. 
This correlation occurs whether we assume MRN or ``meteoritic'' grainsizes.
Since there is a relationship between the optical depth and shell thickness, 
this correlation seems to support models validity.

\begin{figure}[t]
%\resizebox{\hsize}{!}{%
\includegraphics[angle=270,scale=1.0]{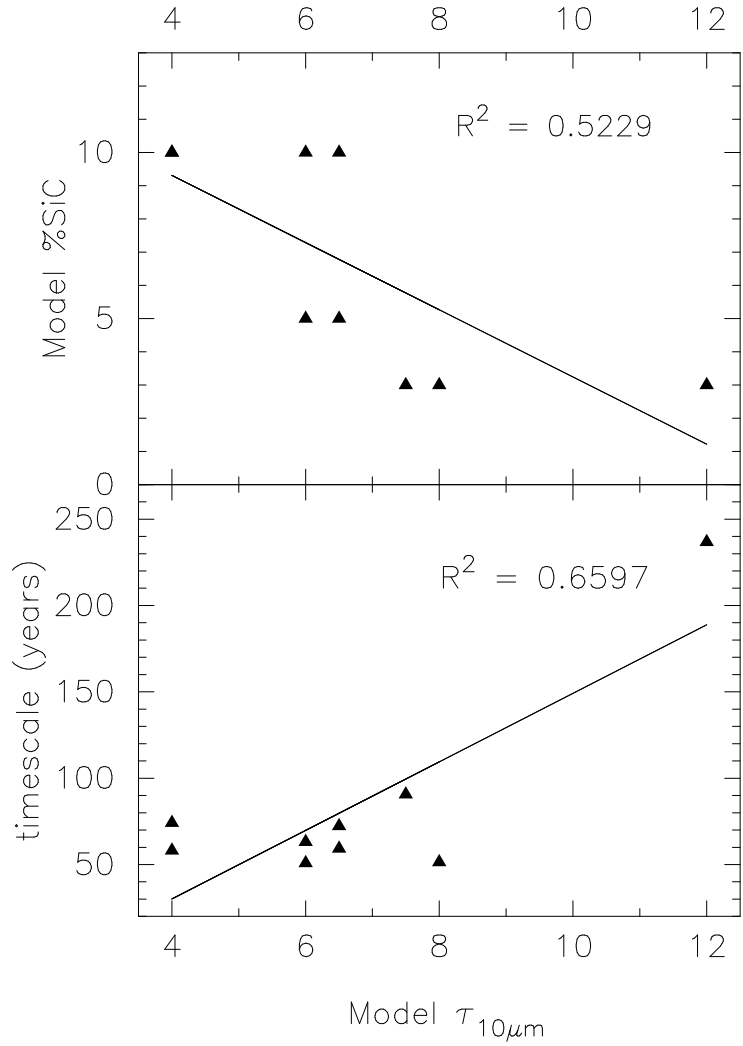}
\caption{\label{correlfigtau} Correlations between the modeled optical depth
and 
({\it top}) the modeled \%age SiC
({\it bottom}) the calculated superwind timescale from Table~\ref{timescaletab}
Solid lines represent the linear regression fit through the points.}
\end{figure}

In addition to this relationship between the modeled optical depth and the 
best-fit blackbody temperature, we found two other parameters that the optical 
depth correlates with: modeled percentage of SiC in the dust shell; and the 
calculated timescale of the obscuring dust. In both cases, these correlations only hold for the generic (MRN) grain-sze distribution models.
(shown in Fig.~\ref{correlfigtau}).
The correlation between optical depth and percentage SiC is such that lower 
optical depths 
require more SiC. This in turn suggests that as mass-loss rates increase, the 
SiC component dwindles. This can be interpreted in two ways:
(1) at these high mass-loss rates SiC gets coated by carbon; and
(2) increased mass-loss rates are associated with higher C/O ratios;
carbon is enriched but not silicon, and thus more carbon 
grains can be made, but not more SiC grains.
The first option seems unlikely in light of meteoritic evidence (\S~\ref{meteor}) and 
theoretical condensation models (\S~\ref{condmod}).
However, it is possible that there is a metallicity effect in play. 
The presolar grains were formed prior to the formation of the solar system, 
and thus their source stars may have had lower metallicities.
\citet{Speck2006,leisenring} argued that coating of SiC grains is more 
likely in lower metallicity environments, and thus it is difficult to see how 
the higher metallicity objects we are now witnessing could have carbon-coated 
SiC grains.
However, \citet{leisenring} argued that SiC grains form the nucleation seed for MgS, 
which may explain this correlation. 
The second option may be logical if outflows are dust driven. 
If more carbon grains can form, the radiation pressure driving the outflow is more effective.

The second optical depth correlation is with the calculated timescale (also 
shown in Fig.~\ref{correlfigtau} and again, only for the MRN grain-size 
distribution). In this case, 
the timescale for the dustshell increases with optical depth. 
The oldest shells have the highest optical depth.
This may simply be due to the pathlength dependence of optical depth 
(see Eqn.~\ref{tau0}), since $\tau$ depends on the geometric size of the dust 
shell the oldest shells should have the largest optical depth. However, this 
should also lead to a correlation between model dust shell thickness and 
optical depth. No such correlation exists.
Another interpretation is that older stars have higher mass-loss rates, and 
thus denser dust shells and higher optical depths. However, this assumes the 
stars all had the same initial mass.
This apparent correlation is not strong, and does not holds for ``meteoritic'' 
grains. Consequently, we will not attempt to place too much importance on it.

There is no correlation between the strength of the 9.7$\mu$m absorption and 
that at 11.3$\mu$m, implying that the carriers of these features may not be 
related. Furthermore, there is no correlation between the 9.7$\mu$m
strength and distance to the object, which leads us to suggest that the cause 
is not interstellar. Obviously there is a strong correlation between the 
strength of the 9.7$\mu$m absorption and the barycentric position of the 
feature, indicating that this is the cause of the broadening.

The most interesting correlations are between the ``isolated'' 
11.3$\mu$m feature (i.e. the residual feature after taking silicate absorption 
into account) and the fitted blackbody temperature. 
There is a correlation both between the SiC 
barycentric position and the blackbody 
temperature and between the  11.3$\mu$m feature-to-continuum ratio 
(i.e. SiC feature strength) and the blackbody temperature (both shown in 
Fig.~\ref{correlfig}). 

The position of the SiC absorption feature moves to shorter wavelengths for 
higher 
blackbody temperatures while also becoming weaker.
This can be understood as being the result of more absorption occuring when 
the surface we see is at a lower temperature, 
which is associated with longer wavelength absorption.

We can interpret this result in terms of the self-absorption scheme described 
by \citet{Speck2005}. The shifts in wavelength were attributed to a 
changes in grain size. However, the weaker absorption and warmer blackbody 
temperatures associated with the shortest wavelength peaks is the opposite 
trend to that described by \citet{Speck2005}. Following their scheme, this 
would imply that the weakest absorption is associated with the largest SiC 
grain. The problem with interpreting this observation is that we do not know 
the C/O ratios for these stars. While C/O does not affect the SiC condensation 
temperature, it will affect the graphite condensation temperature. As seen in 
Fig~\ref{modelfit23166}, we can reduce the inner dust temperature and 
compensate by decreasing both the optical depth and the shell thickness.
This would also change the depth into the shell at which the SiC forms.
It is also possible that the shift in position is due to 
incomplete subtraction of the silicate contribution.
Further studies are ongoing, but are beyond the scope of the 
present work.

The assumption that there is a silicate contribution to the spectrum raises 
several questions. Identifying the source of the 9.7$\mu$m contribution to the
absorption feature remains beyond the scope of the present paper, but, 
needless to say, if it is silicate, the specter of dual chemistry would 
require seriously rethinking our current models of dust formation around 
carbon stars. \citet{lf99} suggested that silicates could form around carbon 
stars, and indeed silicate carbon stars exist.
It is possible that the feature may be attributable to some form 
of hydrogenated amorphous carbon, which has been postulated as the source of 
an $\sim9\mu$m emission feature in optical thin carbon star spectra 
\citep[see][and reference therein]{Thompson2006}.

No additional correlations were found. The lack of correlation between most of 
the observable or modeled parameters echoes 
the results of \citet{Thompson2006}, who found that there are no trends in the 
parameters associated with 11$\mu$m emission feature in visibly observable 
carbon stars. This suggests that even amongst the extreme carbon stars, 
variations in C/O, $s$-process and nitrogen enhancements make discernment of 
the physical properties associated with the 11$\mu$m feature very difficult.

\section{Conclusions}

We have presented three previously unrecognized SiC absorption features in the 
spectra of extreme carbon stars. Together with the seven known SiC absorption 
stars, this brings the total of known extreme carbon stars with SiC absorption 
features to ten.

Previous radiative transfer models of extreme carbon stars utilized relatively 
low condensation temperatures. Here, theoretical condensation models have 
been used to justify much higher condensation temperatures. 
In addition, our models use graphite instead of amorphous carbon, because 
of the preferential formation of graphite at higher temperature and the 
meteoritic evidence. Both the higher condensation temperature (through a 
decrease in the inner radius of the dust shell) and to a smaller extent the 
use of graphite will greatly increase the acceleration felt by the dust grains 
in the shell relative to parameters used in previous research. 

We have shown that grain-size issues cannot be ignored in the production 
of models that accurately fit the observed spectra of extreme carbon stars.  
The size distribution that is needed is not clearly defined because of the 
inherent degeneracy in radiative transfer modeling. Meteoritic grain-size 
distributions are as valid as other size distributions with the advantage of 
being model independent. However, they may underestimate the contribution from 
small grains.

With the exception of IRAS 17534$-$3030, all sample stars could be modeled 
with either the generic (MRN) or ``meteoritic grain-size distribution
IRAS 17534$-$3030's narrow SED required the use of the ``meteoric'' 
distribution. Furthermore, there is no evidence for any molecular absorption 
in its spectrum. Because the 11$\mu$m 
feature is still present in the absence of the other molecular features, it 
supports the attribution of this feature to a solid state carrier.
	
The various parameters compiled in the course of this research (both 
through radiative transfer modeling and from observations) have been compared 
in order to identify any correlations, with the result that the cause of 
differences is the spectra cannot be attributed to mass-loss rate or gas
pressure in the dust condensation zone.
In fact the paucity of correlations between parameters echoes 
the results of \citet{Thompson2006} and 
suggests that even amongst the extreme carbon stars, 
variations in C/O, $s$-process and nitrogen enhancements make 11$\mu$m 
a poor probe of the details of dust shell parameters.

The timescales associated with the heavy mass-loss experienced by these 
extreme carbon stars are very short (tens to hundred of years) and are not
consistent with timescales for the superwind. This indicates that the heavy 
mass-loss phase of carbon stars is not a direct result of thermal pulse 
(although thermal pulses may be the root cause).

\acknowledgments

This work is supported by NSF AST-0607341.
We are very grateful to the referee, Albert Zijlstra, 
for his comments which have significantly improved this paper.
Kevin Volk is also thanked for helpful advice on this paper.




\clearpage
%Fig1

\clearpage
%Fig2

\clearpage
%Fig3

\clearpage
%Fig4

\clearpage
%Fig5

\clearpage
%Fig6

\clearpage
%Fig7

\clearpage
%Fig8

\clearpage

\clearpage

\clearpage

\clearpage

\clearpage

\clearpage

%table1
\begin{table}
\small
\caption{Previous Models of Extreme Carbon stars with 11$\mu$m absorption features \label{prevmod}}
\begin{tabular}{l@{\hspace{1mm}}c@{\hspace{2mm}}c@{\hspace{2mm}}c@{\hspace{1mm}}c@{\hspace{1mm}}c@{\hspace{1mm}}c@{\hspace{1mm}}c@{\hspace{1mm}}c}
\hline
Star& composition$^\dag$& Grain-size &$\tau_{11.3\mu \rm m}$ &	drop-off & R$_{in}^\ast$&  T$_{in}$	& $\dot{M}^\ast$        &	REF\\	
&   (SiC\%)  &  ($\mu$m)             &	                     &	       &($\times 10^{-4}$\,pc)    &	(K)	& M$_\odot$\,yr$^{-1}$	&  \\
\hline
00210+6221 &	0	& 0.1	     & 	4.5	             & 3.00	 & 0.362/D&	1000	& 12$\times 10^{-5}/D$	&	2	\\	
01144+6658 &	8	& 0.1	     &	4.84	             &	         &	  &	1000	& 9.50 $\times 10^{-7}$	&	4	\\	
06582-1507 &	0	& ...        &	5.15	             &	         &	  &	1000	& 1.08$\times 10^{-4}$	&	1	\\	
	   &	0	& 0.1	     &	2.1	             & 2.25      & 1.23   & 1000	& 2.59$\times 10^{-4}$	&	2	\\	
17534$-$3030 &	0	& 0.1	     &	4.4	             & 2.50      & 1.39   & 1000	& 8.20$\times 10^{-4}$	&	2	\\	
19548+3035 &	0	& 0.1	     &	2.5	             & 2.50      & 1.17   & 1000	& 3.89$\times 10^{-4}$	&	2	\\	
21318+5631 &	0	& ...        &	1.36	             &	         &	  &	700	& 1.10$\times 10^{-4}$	&	3	\\	
23166+1655 &	0	& 0.1	     &	1.12	             &	         &	  &	650	& 5.50$\times 10^{-7}$	&	4	\\	
	   &	30	& ...        &	7.85	             &	         &	  &	1000	& 3.30$\times 10^{-5}$	&	1	\\	
	   &	0	& ...        &	1.19	             &	         &	  &	650	& 1.00$\times 10^{-4}$	&	3	\\	
\hline
\end{tabular}
\begin{tabular}{p{5.8in}}
References 1: \citet{Volk1992}; 2: \citet{Volk2000}; 3: \citet{Groenewegen1995}; 4: \citet{Groenewegen1998}\\
$\ast$ originally quoted as a function of distance. Value quoted here assume 
distances from \citet{Groen2002}, listed in Table~\ref{obsparam2}\\
$\dag$ composition assumes remainder dust is carbon. In all but 
\citet{Volk1992} the carbon is amorphous; 
\citet{Volk1992} uses graphitic carbon.\\
Neither \citet{Volk1992} nor \citet{Groenewegen1995} specify the grains sizes 
used in their models\\
\end{tabular}
\end{table}

%table2
\begin{table}
\small
\caption{Target List \label{obstable}}
\begin{tabular}{l@{\hspace{1mm}}l@{\hspace{1mm}}c@{\hspace{1mm}}c@{\hspace{2mm}}c@{\hspace{1mm}}c@{\hspace{1mm}}}
\hline
IRAS &Other & R.A. \footnote{%
Note: Units of right ascension are given in hours, minutes and seconds; 
units of declination are degrees, arcminutes and arcseconds.}& 
Decl.  & TDT  & Date of \\
Number&Names& (J2000)&(J2000) & number & Observation \\
\hline
00210+6221&CGCS 6006& 00 23 51.2& $+$62 38 16.4 &40401901 & 1996 Dec 24\\
01144+6658&V829 Cas, AFGL 190, CGCS 6017&01 17 51.6 &+67 13 55.4&68800128&1997 Oct 03 \\
02408+5458&& 02 44 25.2  & $+$55 11 15& 80002504 & 1998 Jan 24 \\
03313+6058&CGCS 6061&03 35 30.7 &+61 08 47.2&62301907&1997 Jul 31\\
06582+1507&CGCS 6193& 07 01 08.44 & $+$15 03 39.8&71002102 & 1997 Oct 26\\
17534$-$3030&AFGL 5416, CGCS 6690& 17 56 33.1 & $-$30 30 47.1&12102004 & 1996 Mar 17\\
19548+3035&AFGL 2477, CGCS 6851& 21 50 45.0 & $+$53 15 28.0&56100849 & 1997 May 30 \\
21318+5631&AFGL 5625S, CGCS 6888& 21 33 22.98 & $+$56 44 35.0&11101103 & 1997 Mar 7\\
22303+5950&CGCS 6906&22 32 12.8 &+60 06 04.1&77900836&1998 Jan 02\\
23166+1655&LL Peg, AFGL 3068, CGCS 6913& 23 19 12.39 & $+$17 11 35.4&37900867 & 1996 Nov 29\\
\hline
\end{tabular}
\end{table}

%table3
\begin{table}
\caption{Observed parameters of the target sources \label{obsparam}}
\small
%\begin{tabular}{l@{\hspace{2.5mm}}c@{\hspace{0.5mm}}c@{\hspace{0.5mm}}c@{\hspace{0.5mm}}c@{\hspace{0.5mm}}c@{\hspace{0.5mm}}c@{\hspace{0.5mm}}c@{\hspace{0.5mm}}c}
\begin{tabular}{lcccccccc}
\hline
IRAS       &$T_{\rm BB}$& Feature & Feature$^1$  &FWHM & Equivalent & 9.7$\mu$m$^2$ & 11.3$\mu$m$^3$ & SiC$^4$\\ 
\rb{Number}& \rb{(K)}  &\rb{Barycenter}& \rb{Strength} & &\rb{Width} & Strength  & Strength & position\\ 
\hline
00210+6221 & 290K & 10.45 & 0.73 & 2.28&0.60 & 0.76& 0.77 & 11.3\\
01144+6658 & 280K & 10.51 & 0.82 & 2.32&0.35 & 0.88& 0.87 & 11.3\\
02408+5458 & 250K & 11.28 & 0.73 & 2.20&0.30 & --- & 0.83 & 11.3\\
03313+6058 & 350K & 10.12 & 0.87 & 1.72&0.17 & 0.90& 0.97 & 10.8\\
06582+1507 & 320K & 10.25 & 0.91 & 0.70&0.1  & 0.97& 0.94 & 11.0\\
17534$-$3030 & 280K & 10.59 & 0.87 & 0.80&0.30 & 0.92& 0.83 & 11.3\\
19548+3035 & 295K & 10.24 & 0.87 & 2.39&0.25 & 0.88& 0.93 & 11.0\\
21318+5631 & 300K & 10.20 & 0.84 & 2.90&0.37 & 0.88& 0.93 & 11.0\\
22303+5950 & 350K & 10.25 & 0.84 & 1.70&0.25 & 0.86& 0.94 & 11.0\\
23166+1655 & 290K & 11.42 & 0.86 & 2.00&0.21 & --- & 0.88 & 11.4\\
\hline
\end{tabular}
\begin{tabular}{p{6.5in}}
$^1$ Feature strength is the ``peak''-to-continuum ratio and is measured and 
the barycentric position.\\
$^2$ The 9.7$\mu$m strength is the feature-to-continuum ratio measured at 9.7$\mu$m.\\
$^3$ The 11.3$\mu$m strength is the feature-to-continuum ratio measured at 11.3$\mu$m.\\
$^4$ The SiC position, the the approximate barycentric position that the SiC feature 
would have if the short wavelength side of the absorption were due to silicate 9.7$\mu$m 
absorption.\\
\end{tabular}
\end{table}

%table4
\begin{table}
\caption{Compilation of CO mass-loss rates and expansion velocities, together with  distances 
and luminosities of sample stars. \label{obsparam2}}
\begin{tabular}{lcccccc}
\hline
Source Name & $v_{\rm exp}$ &	D & $\dot{M}_{\rm gas}$ & $\dot{M}_{\rm dust}$ & L$_{\star}$ \\
	    &	km/s	    & kpc & M$_\odot$\,yr$^{-1}$& M$_\odot$\,yr$^{-1}$ & L$_\odot$ \\
\hline
00210+6221* &   16.7        & 3.97& 3.02$\times 10^{-5}$& ---                 &   1$\times 10^{4}$  \\   
01144+6658  &	18	    & 2.78& 6.38$\times 10^{-5}$&1.51$\times 10^{-7}$ &1.69$\times 10^{4}$\\
02408+5458* &	11	    & 5.3 & 1.60$\times 10^{-5}$&3.5 $\times 10^{-7}$ &5.70$\times 10^{3}$\\
03313+6058  &	13.9	    & 5.24& 2.37$\times 10^{-5}$&1.14$\times 10^{-7}$ &1.31$\times 10^{4}$\\
06582+1507  &	13.7	    & 4.7 & 1.43$\times 10^{-5}$&1.07$\times 10^{-7}$ &1.32$\times 10^{4}$\\
17534$-$3030  &	19	    & 2	  & 3.76$\times 10^{-5}$&1.06$\times 10^{-7}$ &1.21$\times 10^{4}$\\
19548+3035  &	22.3	    & 3.38& 1.14$\times 10^{-4}$&2.15$\times 10^{-7}$ &1.32$\times 10^{4}$\\
21318+5631  &	19.6	    & 1.77& 7.69$\times 10^{-6}$&1.18$\times 10^{-7}$ &1.24$\times 10^{4}$\\
22303+5950  &	18.3	    & 3.86& 3.19$\times 10^{-4}$&1.09$\times 10^{-7}$ &1.25$\times 10^{4}$\\
23166+1655  &	15.1	    & 1	  & 1.44$\times 10^{-5}$&8.27$\times 10^{-8}$ &1.10$\times 10^{4}$\\
\hline
\end{tabular}
\begin{tabular}{p{5.5in}}
All data are from \citet{Groen2002} except those marked with $\ast$. 
IRAS\,02408+5458 data come from \citet{Groen1999}.
IRAS\,00210+6221 data are compiled from 
\citet{Volk2000} \citep[$\dot{M}$, which also convolves the distance from ][]{Groen2002}; and
\citet{Volk1992} ($v_{\rm exp}$).
\end{tabular}
\end{table}

%table 5
\begin{table}
\caption{Model and Derived Parameters \label{tabfit1}}
%\begin{tabular}{lcccp{1.85cm}@{\hspace{7.5mm}}r}
\begin{tabular}{lrrrrcc}
\hline
IRAS  & T$_\star$ &T$_{\rm inner}$&$R_{\rm out}/R_{\rm in}$ & $\tau_{10\mu \rm m}$ &\multicolumn{2}{c}{Composition}\\
Number&           &               &                         &                      &  SiC\% & Graphite \% \\
\hline
\multicolumn{7}{c}{MRN grain size distribution}	\\
00210+6221	&	3000	&	1800	&	10	&	6	&	10	&	90	\\
01144+6658	&	3000	&	1800	&	15	&	6.5	&	10	&	90	\\
02408+5458	&	3000	&	1800	&	20	&	12	&	3	&	97	\\
03313+6058	&	3000	&	1800	&	15	&	4	&	10	&	90	\\
06582+1507	&	3000	&	1800	&	10	&	6	&	5	&	95	\\
17534-3030	&	3000	&	1800	&		&		&		&		\\
19548+3035	&	3000	&	1800	&	15	&	6.5	&	5	&	95	\\
21318+5631	&	3000	&	1800	&	10	&	8	&	3	&	97	\\
22303+5950	&	3000	&	1800	&	15	&	4	&	10	&	90	\\
23166+1655	&	3000	&	1800	&	15	&	7.5	&	3	&	97	\\
\hline
\multicolumn{7}{c}{Meteoritic grain size distribution}	\\
00210+6221	&	3000	&	1800	&	10	&	6	&	30	&	70	\\
01144+6658	&	3000	&	1800	&	10	&	7.5	&	30	&	70	\\
02408+5458	&	3000	&	1800	&	100	&	8.5	&	12	&	88	\\
03313+6058	&	3000	&	1800	&	500	&	3.25	&	40	&	60	\\
06582+1507	&	3000	&	1800	&	10	&	6	&	20	&	80	\\
17534-3030	&	3000	&	1800	&	20	&	5	&	35	&	65	\\
19548+3035	&	3000	&	1800	&	100$^\ast$&	4.75	&	25	&	75	\\
21318+5631	&	3000	&	1800	&	100	&	5	&	20	&	80	\\
22303+5950	&	3000	&	1800	&	500	&	3.25	&	40	&	60	\\
23166+1655	&	3000	&	1800	&	50	&	5	&	25	&	75	\\
\hline
\end{tabular}
\begin{tabular}{p{16.8cm}}
$\ast$ Models with $R_{\rm out}/R_{\rm in} > 100$ can be accommodated because the data beyond 26$\mu$m is poor and ignored, but based on 
the similarity of this source to IRAS\,21318+5631, we assume this is a good upper limit.\\
\end{tabular}
\end{table}

\begin{table}
\caption{Size and age of the dust shells \label{timescaletab}}
\begin{tabular}{lcccccc}
\hline
        & \multicolumn{3}{c}{MRN size distribution}  & \multicolumn{3}{c}{Meteoritic size distribution}\\
IRAS    & R$_{\rm in}$ &R$_{\rm out}$ & Age    & R$_{\rm in}$ &R$_{\rm out}$ & Age    \\
Number  & ($10^{14}$cm)&($10^{14}$cm) & (yrs)  & ($10^{14}$cm)&($10^{14}$cm) & (yrs)  \\
\hline
00210	& 2.68	&	26.8	&       50.9  &	2.78	& 	2.78	&	52.8	\\
01144	& 2.74	&	41.1	&	72.4  &	3.08	& 	30.8	&	54.2	\\
02408	& 4.11	&	82.2	&	236.8 &	2.99	& 	299	&	861.3	\\
03313	& 2.17	&	32.6	&	74.2  &	2.10	& 	1050	&	2393.7	\\
06582	& 2.73	&	27.3	&	63.1  &	2.72	& 	27.2	&	62.9	\\
17534	& 2.51	&	50.2	&	83.7  &	2.51	& 	50.2	&	83.7	\\
19548	& 2.78	&	41.7	&	59.3  &	2.38	& 	238 	&	338.2	\\
21318	& 3.18	&	31.8	&	51.4  &	2.41	& 	241	&	389.6	\\
22303	& 2.24	&	33.6	&	58.2  &	2.10	& 	1050	&	1818.2	\\
23166	& 2.88	&	43.2	&	90.7  &	2.45	& 	123	&	257.1	\\
\hline
\end{tabular}
\begin{tabular}{p{6in}}
Calculation of R$_{\rm out}$ is based on the model parameters listed in Table~\ref{tabfit1}. R$_{\rm in}$ comes from the model output. The calculation of the age of the dust shells is then done using these data and the observed expansion velocities listed in Table~\ref{obsparam2}
\end{tabular}
\end{table}

\end{document}